\documentclass[12pt]{article}
\usepackage{epsf,amsmath,euscript}

\def\pslash{\rlap{\hspace{0.02cm}/}{p}}
\def\delslash{\rlap{\hspace{0.02cm}/}{\partial}}
\def\lslash{\rlap{/}{l}}

\def\nslash{\rlap{\hspace{0.02cm}/}{n}}
\def\nbslash{\rlap{\hspace{0.02cm}/}{\bar n}}
\def\vslash{\rlap{\hspace{0.02cm}/}{v}}

\def\Dslash{\rlap{\hspace{0.07cm}/}{D}}
\def\Aslash{\rlap{\hspace{0.07cm}/}{A}}
\def\calAslash{\rlap{\hspace{0.08cm}/}{{\EuScript A}}}
\def\calDslash{\rlap{\hspace{0.1cm}/}{{\EuScript D}}}

\def\A{{\EuScript A}}
\def\D{{\EuScript D}}
\def\H{{\EuScript H}}
\def\Q{{\EuScript Q}}
\def\X{{\EuScript X}}

\begin{document}

\begin{titlepage}

\begin{flushright}
CLNS~02/1803\\
SLAC-PUB-9549\\
{\tt hep-ph/0211018}\\[0.2cm]
November 1, 2002
\end{flushright}

\vspace{0.7cm}
\begin{center}
\Large\bf 
Spectator Interactions in Soft-Collinear\\ 
Effective Theory
\end{center}

\vspace{0.8cm}
\begin{center}
{\sc Richard J.~Hill$^a$ and Matthias Neubert$^b$}\\
\vspace{0.7cm}
{\sl ${}^a$Stanford Linear Accelerator Center, Stanford University\\
Stanford, CA 94309, U.S.A.\\
\vspace{0.3cm}
${}^b$Newman Laboratory for Elementary-Particle Physics, Cornell 
University\\
Ithaca, NY 14853, U.S.A.}
\end{center}

\vspace{1.0cm}
\begin{abstract}
\vspace{0.2cm}\noindent 
Soft-collinear effective theory is generalized to include soft massless 
quarks in addition to collinear fields. This extension is necessary for
the treatment of interactions with the soft spectator quark in a heavy 
meson. The power counting of the relevant fields and the construction of 
the effective Lagrangian are discussed at leading order in $\Lambda/m_b$. 
Several novel effects occur in the matching of full-theory amplitudes 
onto effective-theory operators containing soft light quarks, such as the 
appearance of an intermediate mass scale and large non-localities of 
operators on scales of order $1/\Lambda$. Important examples of 
effective-theory operators with soft light quarks are studied and their 
renormalization properties explored. The formalism presented here forms 
the basis for a systematic analysis of factorization and power 
corrections for any exclusive $B$-meson decay into light particles.
\end{abstract}
\vfil

\end{titlepage}

\section{Introduction}

Processes involving energetic light particles play an important role in
particle physics. Examples are jet production in $e^+ e^-$ annihilation,
$B$-meson decays into light particles, and many other hard QCD processes. 
The theoretical description of such processes is often complicated by the 
presence of soft and collinear singularities, which invalidate the 
application of the (local) operator product expansion. In some cases 
factorization theorems have been established, which provide a simplified 
description of the relevant observables at leading order in the limit 
$E\gg\Lambda$, where $E$ is the characteristic energy of the process, and 
$\Lambda\sim\Lambda_{\rm QCD}$ is the scale of non-perturbative hadronic 
physics. Formal proofs of these factorization theorems are difficult and
typically rely on a diagrammatic analysis of different momentum regions 
giving rise to leading-order contributions to the amplitude. It would be
desirable to facilitate these proofs and make them more transparent. Even 
more challenging is to develop a systematic framework for the 
parameterization and classification of power corrections for observables 
that do not admit an expansion in local operators. 

The proposal of an effective field theory for collinear and soft 
particles by Bauer et al.\ 
\cite{Bauer:2000ew,Bauer:2000yr,Bauer:2001ct,Bauer:2001yt} is an 
important step toward achieving this goal. This ``soft-collinear 
effective theory'' (SCET) lets us discuss factorization theorems and 
power corrections in terms of fields and operators rather than momentum 
regions of Feynman diagrams. While power counting in the effective theory 
is non-trivial due to the fact that the relevant operators are non-local, 
there is nevertheless hope for a controlled expansion of amplitudes in 
terms of hadronic matrix elements of SCET operators (multiplied by 
perturbative Wilson coefficient functions), whose structure is 
constrained by gauge invariance and Lorentz symmetry.

SCET has been applied to prove factorization and resum Sudakov logarithms 
for the endpoint region of the photon energy spectrum in $B\to X_s\gamma$ 
decays \cite{Bauer:2000ew,Bauer:2000yr}, and to prove QCD factorization 
as established in \cite{Beneke:2000ry} for the weak decays $B\to D\pi$ 
\cite{Bauer:2001cu}. Applications to hard processes outside $B$ physics, 
such as deep-inelastic scattering, Drell--Yan production, and deeply 
virtual Compton scattering, have been considered in \cite{Bauer:2002nz}.
Recently, the formulation of SCET has been extended beyond leading power 
\cite{Chay:2002vy,Beneke:2002ph}. These analyses constitute a significant
first step toward a theory of power corrections that is more general than 
a local operator product expansion.

In this work we present an extension of the previous formulation of SCET, 
which is necessary for the discussion of exclusive $B$ decays into light 
particles, such as $B\to\pi\pi$, $B\to K^*\gamma$, and heavy-to-light 
form factors at large recoil. (The ``soft contribution'' to 
heavy-to-light form factors has been discussed previously in the context 
of SCET, both at leading order \cite{Bauer:2000yr} and beyond 
\cite{Chay:2002vy,Beneke:2002ph}. However, no systematic treatment of all 
leading-power contributions has been presented so far.) For all these 
cases QCD factorization theorems have been proposed 
\cite{Beneke:1999br,Beneke:2001ev,Bosch:2001gv,Beneke:2001at,Beneke:2000wa},
but have not yet been proved beyond next-to-leading order in perturbation
theory. In fact, factorization theorems for exclusive $B$-meson decays 
into light particles are more complicated than those for the decays into 
a heavy-light final state such as $B\to D\pi$. In addition to a 
form-factor term, a hard-scattering contribution appears at leading 
power, which results from hard gluon exchange with the spectator quark in 
the $B$ meson \cite{Beneke:1999br,Beneke:2001ev}. This contribution 
involves a convolution of a hard-scattering kernel with light-cone 
distribution amplitudes for the final-state hadrons and the initial $B$ 
meson. A proof of factorization for such spectator contributions has not 
been attempted so far. The formulation of SCET developed here provides 
for the first time the framework for a systematic discussion of 
factorization in all of these cases and others 
\cite{Korchemsky:1999qb,Bosch:2002bv,Descotes-Genon:2002mw}. 

Whereas for highly energetic light mesons the relevance of light-cone 
distribution amplitudes to the description of exclusive processes is 
familiar from many applications of perturbative QCD, relatively little is 
known about the light-cone structure of heavy hadrons. At first sight, 
even the appearance of light-cone distributions for the $B$ meson seems 
surprising, because (in the $B$-meson rest frame) all characteristic
momentum scales are soft, of order $\Lambda$. Unlike for a fast light 
meson, there is thus no hierarchy between the different components of the 
momenta of the $B$-meson constituents. However, the kinematics of 
heavy-to-light decay processes ensures that (in some cases) only the 
projection of the soft spectator momentum along some light-like direction 
enters the decay amplitudes at leading power in $\Lambda/m_b$. Because of 
the softness of the relevant momentum scales the notion of twist is not 
appropriate for the characterization of $B$-meson light-cone distribution 
amplitudes, which instead should be categorized according to their 
canonical dimension. Some distinctive features between $B$-meson and 
light-meson distribution amplitudes have already been noted in the 
literature. Whereas there exists a single leading-twist distribution 
amplitude for light pseudoscalar and vector mesons, two independent 
$B$-meson distribution amplitudes appear at leading order in the 
heavy-quark expansion \cite{Grozin:1996pq}. In the case of light mesons, 
the equations of motion imply relations between higher-twist distribution 
amplitudes and connect them with amplitudes of lower twist (see, e.g., 
\cite{braun,Ball:1998sk}). For heavy mesons instead, the equations of 
motion relate the leading-order two-particle amplitudes to certain 
three-particle amplitudes (corresponding to quark--antiquark--gluon Fock 
components) of higher dimension \cite{Beneke:2000wa,Kawamura:2001jm}.

The intrinsic softness of the $B$-meson internal dynamics complicates the 
understanding of factorization properties of decay amplitudes. A new 
element is the appearance of an intermediate scale of order $m_b\Lambda$, 
which arises from the scalar product of a soft spectator momentum $l$ 
with a collinear momentum $p_c$. While this scale is perturbative (since 
formally $p_c\cdot l\gg\Lambda^2$) and thus should be integrated out from 
the low-energy effective theory, it nevertheless depends on the $B$-meson 
dynamics and is not simply fixed by kinematics. This is different from 
previous applications of heavy-quark effective theory (HQET) and SCET, 
where the large scales were fixed in terms of the $b$-quark mass and the 
large energy $E\sim m_b$ carried by collinear fields. The appearance of 
an intermediate scale naturally leads to non-local operators integrated 
along light-like directions, whose matrix elements define the $B$-meson 
distribution amplitudes. The presence of three widely separated scales 
($m_b^2\gg m_b\Lambda\gg\Lambda^2$) also complicates the perturbative 
structure of decay amplitudes. Sudakov double logarithms appear at every 
order in perturbation theory and must be resummed. 

The remainder of this paper is organized as follows: In 
Section~\ref{sec:SCET} we present the construction of the SCET relevant 
to exclusive $B$ decays, introduce the relevant fields, discuss their 
power counting and gauge transformations, and derive the effective 
Lagrangian. Interactions between soft and collinear fields are studied in 
Section~\ref{sec:SCints}, and are shown to be absent at leading order. In 
Section~\ref{sec:softcol} we discuss in detail the matching of current 
operators containing a soft light quark and a collinear quark onto 
operators in the effective theory. Several new features appear in this 
calculation, such as the emergence of the intermediate scale, large 
non-localities of operators on a scale $1/\Lambda$, and unsuppressed
couplings between soft quarks and transverse collinear gluons. In 
Section~\ref{sec:RPI} we show how reparameterization invariance can be 
used to constrain the functional dependence of short-distance coefficient 
functions on the separation between the component fields of non-local 
operators. The renormalization of such operators and the related 
resummation of Sudakov logarithms are briefly discussed. The matching of 
local four-quark operators onto operators in the SCET is studied in 
Section~\ref{sec:4qops}. This application is of relevance to QCD 
factorization theorems for many exclusive $B$ decays. A surprising result 
is that, generically, higher-twist three-particle distribution amplitudes 
of a light final-state meson can contribute to the decay amplitude at 
leading power. In Section~\ref{sec:examples} we illustrate the 
implications of these findings for the factorization properties of 
$B$-meson decay amplitudes in some toy models. We present two examples, 
one where a standard QCD factorization formula can be established at 
leading power, and one where the factorization formula must be 
generalized due to non-trivial interactions of the soft spectator quark 
with collinear gluons. The results derived in this work suggest a new 
formulation of the SCET, in which operators are composed out of 
gauge-invariant building blocks, replacing the original quark and gluon 
fields. This formalism is developed in Section~\ref{sec:pretty}. In the 
new formulation operators are automatically gauge invariant and their 
structure is constrained only by Lorentz invariance. Finally, in 
Section~\ref{sec:formfactor} we explain how our power-counting scheme can 
be applied to describe the soft overlap contribution to heavy-to-light 
form factors, which in previous work was analyzed using a different 
formulation of the SCET \cite{Chay:2002vy,Beneke:2002ph}. A summary of 
our findings is given in Section~\ref{sec:concl}.

\section{Ingredients of the effective theory}
\label{sec:SCET}

We start by discussing in detail the properties of the fields present in 
the low-energy theory, using the coordinate-space formulation developed 
by Beneke et al.\ \cite{Beneke:2002ph} (and avoiding the hybrid
momentum--position space representation and label-operator formalism 
employed in earlier papers on SCET). Part of this discussion is a 
repetition of similar results presented in that paper and earlier work 
(see, in particular, the review \cite{Bauer:2001yt}), but this will be 
necessary in order to set up our notations in a self-contained way. Since 
our power counting and choice of degrees of freedom is different from the 
one employed in \cite{Beneke:2002ph}, our results for the effective 
Lagrangian and external operators in SCET will be different from those 
already discussed in that work. 

Our focus in this paper is on exclusive $B$-meson decays into final 
states containing light, energetic particles. The invariant masses of 
the final-state hadrons are of order $\Lambda$, and the momenta of their 
constituents are predominantly collinear. It is often convenient to 
decompose momenta and gauge fields in the light-cone basis constructed 
with the help of two light-like vectors
\begin{equation}
   n^\mu = (1,0,0,1) \,, \qquad \bar n^\mu = (1,0,0,-1) \,,
\end{equation}
which obey $n^2=\bar n^2=0$, and $n\cdot\bar n=2$. An arbitrary 4-vector
can be expanded as $p^\mu=\frac12(p_+\bar n^\mu+p_- n^\mu)+p_\perp^\mu$ 
with $p_+=n\cdot p$ and $p_-=\bar n\cdot p$. The components 
$(p_+,p_-,p_\perp)$ of a collinear momentum scale like 
$p_c\sim E(\lambda^2,1,\lambda)$, where $E\sim m_b$ is the large energy 
release in the process, and $\lambda\sim\Lambda/E$ is the expansion 
parameter of the SCET. The initial $B$ meson, on the other hand, consists 
of soft partons with momenta scaling like 
$p_s\sim E(\lambda,\lambda,\lambda)$. (This assumes that we work in the 
$B$-meson rest frame and subtract the static piece $m_b\,v^\mu$ from the 
$b$-quark momentum.) Note that in kinematical situations different from 
the ones considered here the scaling relations of soft and collinear 
momenta (and of the corresponding SCET fields) can be different. For 
instance, when the photon energy in inclusive $B\to X_s\gamma$ decays is 
near the kinematic endpoint, the hadronic final state $X_s$ has an 
invariant mass of order $\sqrt{E\Lambda}$, which is much larger than 
$\Lambda$. It is then appropriate to introduce an expansion parameter 
$\lambda\sim\sqrt{\Lambda/E}$ \cite{Bauer:2000yr}. 

In the kinematical situation of relevance to our discussion, the fields 
appearing in the low-energy effective theory are either soft or 
collinear. As in HQET, we introduce a soft heavy-quark field $h_v$ 
defined in terms of the QCD field $b$ by
\begin{equation}
   h_v(x) = e^{im_b v\cdot x}\,\frac{1+\vslash}{2}\,b(x) \,, \qquad
   \mbox{with~~} \vslash\,h_v = h_v \,,
\end{equation}
where $v$ is the $B$-meson velocity. The Fourier modes of the field $h_v$ 
carry the soft residual momentum $k^\mu=p_b^\mu-m_b\,v^\mu$. The two 
components of the spinor $b$ projected out in the definition of $h_v$ are 
integrated out in the construction of the HQET. The soft light-quark 
field $q_s$ and soft gluon fields $A_s^\mu$ are simply given by the 
corresponding QCD fields, restricted to the subspace of soft Fourier 
modes. Using the identity $n\cdot\bar n=2$, the Dirac field $\psi_c$ of a 
collinear quark can be decomposed into two 2-component spinors
\begin{equation}
   \xi_n = \frac{\nslash\,\nbslash}{4}\,\psi_c \,, \qquad
   \eta_n = \frac{\nbslash\,\nslash}{4}\,\psi_c \,, \qquad
   \mbox{with~~} \nslash\,\xi_n = \nbslash\,\eta_n = 0 \,.
\end{equation}
The components of $\eta_n$ are suppressed with respect to those of 
$\xi_n$ by a factor $\lambda\sim\Lambda/E$ and are integrated out in the 
construction of the SCET. The collinear gluon fields $A_{c,n}^\mu$ are 
defined as in QCD, restricted however to the subspace of Fourier modes 
with collinear momenta. For the sake of simplicity, we will from now on 
drop the label $v$ on heavy-quark fields, and the label $n$ on collinear 
quark and gluon fields. It is understood that collinear fields always 
have their large momentum component in the $n$-direction, with 
$\bar n\cdot p_c>0$.

From the scaling behavior of the various two-point functions of two soft
or two collinear fields one can derive the scaling properties of these
fields with the expansion parameter $\lambda$. The strategy here is to
define the kinetic terms in the action to have scaling $\lambda^0$, so 
that factors of $\lambda$ only appear in vertices of the effective 
theory. It follows that the soft fields scale like
$h,\,q_s\sim\lambda^{3/2}$ and $A_s^\mu\sim(\lambda,\lambda,\lambda)$, 
whereas the collinear fields scale like $\xi\sim\lambda$, 
$\eta\sim\lambda^2$, and $A_c^\mu\sim(\lambda^2,1,\lambda)$. Note that 
the covariant derivatives $iD_s^\mu\equiv i\partial^\mu+g A_s^\mu$ and 
$iD_c^\mu\equiv i\partial^\mu+g A_c^\mu$ have homogeneous scaling laws 
when acting on soft or collinear fields, respectively.

Previous discussions of SCET have often introduced ultrasoft modes with 
momentum scaling $p_{us}\sim E(\lambda^2,\lambda^2,\lambda^2)$, because
these modes can be coupled to collinear particles without taking them far 
off their mass shell. In cases where the expansion parameter scales like
$\lambda\sim\sqrt{\Lambda/E}$ ultrasoft fields simply correspond to what
we call soft modes in the present paper. In our case, where 
$\lambda\sim\Lambda/E$, there is no need to introduce ultrasoft modes as 
degrees of freedom in the effective theory, because there are no external 
ultrasoft particles present. Such modes would correspond to color fields 
extending over large distance scales of order $m_b/\Lambda^2$, which do 
not appear in QCD because of confinement.

Operators in the effective theory must be invariant under residual gauge 
transformations in the collinear and soft sectors (i.e., transformations 
that leave the scaling properties of the fields unaltered). Under a 
collinear gauge transformation $U_c(x)$ the collinear fields transform 
according to $\xi\to U_c\,\xi$ and $A_c^\mu\to U_c\,A_c^\mu\,U_c^\dagger
+(i/g)\,U_c\,(\partial^\mu U_c^\dagger)$, whereas soft fields remain 
invariant. Likewise, under a soft gauge transformation $U_s(x)$ we have 
$h\to U_s\,h$, $q_s\to U_s\,q_s$, and $A_s^\mu\to U_s\,A_s^\mu\,
U_s^\dagger+(i/g)\,U_s\,(\partial^\mu U_s^\dagger)$, while collinear 
fields remain invariant. Gauge invariance of operators built out of these 
fields can be restored by the introduction of collinear and soft Wilson 
lines defined as
\begin{equation}\label{WSdef}
\begin{aligned}
   W(x) &= P\exp\left( ig\int_{-\infty}^0\!ds\,
    \bar n\cdot A_c(x+s\bar n) \right) , \\
   S(x) &= P\exp\left( ig\int_{-\infty}^0\!dt\,
    n\cdot A_s(x+tn) \right) ,
\end{aligned}
\end{equation}
which can be visualized as color strings attaching to a quark field at 
point $x$ and extending to infinity. (It does not matter whether the 
integrals in these expressions run from $-\infty$ to 0 or from 0 to 
$+\infty$.) The path-ordering symbol ``$P$'' is defined such that the 
gluon fields are ordered from left to right in order of decreasing $s$ or 
$t$ values. We also need the conjugate operators $W^\dagger(x)$ and 
$S^\dagger(x)$, which are given by analogous expressions with $ig$ 
replaced by $-ig$, and with the opposite ordering of the fields. Under a 
collinear gauge transformation 
$W(x)\to U_c(x)\,W(x)\,U_c^\dagger(-\infty)$, whereas $S(x)$ remains 
invariant. Similarly, under a soft gauge transformation 
$S(x)\to U_s(x)\,S(x)\,U_s^\dagger(-\infty)$, whereas $W(x)$ remains 
invariant. If, without loss of generality, we agree that the fields do 
not transform at infinity, then it follows that $W^\dagger(x)\,\xi(x)$ is 
gauge invariant, as are $S^\dagger(x)\,h(x)$ and $S^\dagger(x)\,q_s(x)$.
The Wilson lines satisfy several important properties, the most useful
ones being
\begin{equation}\label{neat}
   W^\dagger\,i\bar n\cdot D_c\,W = i\bar n\cdot\partial \,, \qquad
   \frac{1}{i\bar n\cdot D_c+i\epsilon}
   = W\,\frac{1}{i\bar n\cdot\partial+i\epsilon}\,W^\dagger \,,
\end{equation}
and corresponding relations for $S$. Also note that 
$(i\bar n\cdot D_c\,W)=0$ and $(in\cdot D_s\,S)=0$ by definition.

A particularly important object in our discussion below is the 
combination $\A_c^\mu=W^\dagger(iD_c^\mu\,W)$, which is invariant under 
both collinear and soft gauge transformations. By definition 
$\bar n\cdot\A_c=0$, but the other components of $\A_c^\mu$ are 
non-zero. In the light-cone gauge $\bar n\cdot A_c=0$, we have $W=1$ 
and hence $\A_c^\mu=g A_c^\mu$. In an arbitrary gauge, we find the 
useful representation
\begin{equation}\label{Adef}
   \A_c^\mu(x) = \big[ W^\dagger(iD_c^\mu\,W) \big](x)
   = \int_{-\infty}^0\!dw\,\bar n_\alpha\,\big[ W^\dagger 
   g G_c^{\alpha\mu}\,W \big](x+w\bar n) \,,
\end{equation}
which makes explicit that $\A_c^\mu$ is a pure color octet,
$\A_c^\mu=\A_c^{\mu,a}\,t_a$. (Here and below color indices on gluon 
fields will appear as superscripts, while subscripts ``$c$'' and 
``$s$'' always refer to ``collinear'' or ``soft'', respectively.) To 
derive this formula one notes that both sides are gauge invariant, and 
that the result is obviously correct in the light-cone gauge. The reader 
should think of the object $\A_c$ as an insertion of a collinear gluon 
field, remembering however that this quantity is gauge invariant.

The effective Lagrangian for soft and collinear fields can be derived by
systematically integrating out the hard modes (including the 
small-component fields for heavy and collinear quarks) from the QCD 
Lagrangian, and expanding the result in powers of $\lambda$. It is 
convenient to split up the answer into several terms, 
\begin{equation}\label{Leff}
   {\cal L}_{\rm SCET} = {\cal L}_h + {\cal L}_s + {\cal L}_c
   + {\cal L}_g + {\cal L}_{sc} \,.
\end{equation}
The effective Lagrangian for heavy quarks is the familiar HQET Lagrangian 
\cite{Neubert:1993mb}
\begin{equation}\label{HQET}
   {\cal L}_h = \bar h\,iv\cdot D_s\,h
   + \frac{1}{2m_b}\,\bigg[ \bar h\,(iD_s)^2\,h
   + C_{\rm mag}(\mu)\,\bar h\,\frac{g}{2}\,\sigma_{\mu\nu}\,
   G_s^{\mu\nu}\,h \bigg] + O(1/m_b^2) \,.
\end{equation}
Let us note parenthetically that interactions between heavy quarks and 
soft gluons are, strictly speaking, not allowed in the low-energy theory, 
since they put the heavy quark off-shell by an amount 
$(m_b v+k)^2-m_b^2\sim E\Lambda$. These interactions can be integrated 
out, leading to the replacement $h\to S_v\,h_0$, where $S_v$ is a Wilson 
line defined in analogy with (\ref{WSdef}) but with $n$ replaced by the 
$B$-meson velocity $v$. The field $h_0$ is sterile in the sense that it 
does not couple to soft or collinear gluons. The heavy-quark Lagrangian 
is then simply ${\cal L}_h=\bar h_0\,iv\cdot\partial\,h_0+O(1/m_b)$. In 
all expressions for operators containing heavy-quark fields the 
replacement $h\to S_v\,h_0$ must be made as well. Since this has no 
effect on the Feynman rule for the soft-gluon couplings, rather than 
introducing the new string operator $S_v$ we will follow the usual 
convention of using the HQET field $h$ with its soft interactions as 
given by (\ref{HQET}).

The Lagrangian for a soft massless quark is the usual Dirac Lagrangian 
(it would be straightforward to include a small mass term)
\begin{equation}\label{Lsoft}
  {\cal L}_s = \bar q_s\,i\Dslash_s\,q_s \,.
\end{equation}
The effective Lagrangian for collinear quarks, which is obtained by 
integrating out the small-component field $\eta$, has a more interesting 
structure \cite{Bauer:2000yr}. With our notations it reads 
\begin{eqnarray}\label{Lcol}
   {\cal L}_c(x) 
   &=& \bar\xi\,\frac{\nbslash}{2}\,in\cdot D_c\,\xi 
    + \bar\xi\,\frac{\nbslash}{2}\,i\Dslash_{c\perp}\,
    \frac{1}{i\bar n\cdot D_c+i\epsilon}\,i\Dslash_{c\perp}\,\xi
    \nonumber\\
   &=& \bar\xi(x)\,\frac{\nbslash}{2}\,in\cdot D_c\,\xi(x) 
    - i\int_{-\infty}^0\!ds\,\big[
    \bar\xi\,i\overleftarrow{\Dslash}_{c\perp} W \big](x)\,
    \frac{\nbslash}{2}\,\big[ W^\dagger i\Dslash_{c\perp}\,\xi
    \big](x+s\bar n) \,.
\end{eqnarray}
In \cite{Beneke:2002ph} it has been argued that the choice of the 
$+i\epsilon$ prescription is arbitrary and not dictated by the QCD 
Lagrangian. A regularization of the inverse differential operator is 
necessary but bears no physical implications. The Lagrangian ${\cal L}_c$ 
sums up an infinite number of leading-order couplings between collinear 
quarks and (scalar or longitudinal) gluons. In the absence of sources the 
collinear Lagrangian is related to the QCD Lagrangian by a Lorentz boost, 
and so the two must be equivalent. As a result, the collinear Lagrangian 
is exact to all orders in $\lambda$, and it is not renormalized 
\cite{Beneke:2002ph}. Finally, the pure-glue Lagrangian ${\cal L}_g$ in 
(\ref{Leff}) takes the same form as in QCD, including gauge-fixing and 
ghost terms. However, it is understood that no term in this Lagrangian 
couples soft to collinear gluon fields. Those couplings, if present, 
would be part of ${\cal L}_{sc}$.

\section{Soft-collinear interactions}
\label{sec:SCints}

As explained in \cite{Beneke:2002ph}, there are no vertices in the 
effective theory that connect the two heavy-quark fields to any number of 
collinear fields, because such interactions are kinematically forbidden 
at tree level. By the Coleman--Norton theorem \cite{Coleman} they can 
therefore not lead to on-shell singularities in any QCD diagram, and 
hence there is no need to include such interactions as part of the 
effective Lagrangian. However, the same argument does not apply to the 
interactions between collinear particles and soft light fields. We will 
now investigate these interactions in more detail.

It is kinematically allowed to couple soft fields to collinear fields 
only if the total soft momentum $p_s^{\rm tot}$ satisfies the condition 
$n\cdot p_s^{\rm tot}=O(E\lambda^2)$, and the difference between the 
ingoing and outgoing collinear momenta is such that 
$\bar n\cdot(p_c^{\rm out}-p_c^{\rm in})=O(E\lambda)$. These constraints 
imply a power counting for the measure $d^4x$ in the soft-collinear 
action $\int d^4x\,{\cal L}_{sc}$ that is different from that for the 
interactions of soft or collinear fields among themselves. Usually the 
measure $d^4 x$ is counted as $\lambda^{-4}$, since it eliminates either 
a collinear or a soft momentum, with $d^4p_c\sim d^4p_s\sim\lambda^4$. 
As a result, the operators appearing at leading order in the Lagrangians 
${\cal L}_h$, ${\cal L}_s$, ${\cal L}_c$ and ${\cal L}_g$ scale like 
$\lambda^4$. On the other hand, when a term in the soft-collinear 
Lagrangian ${\cal L}_{sc}$ is integrated over $d^4x$, this produces 
$\delta$-functions
\begin{equation}\label{delta}
   \delta(\bar n\cdot p_c^{\rm out}-\bar n\cdot p_c^{\rm in}
          -\bar n\cdot p_s^{\rm tot})\,
   \delta(n\cdot p_c^{\rm out}-n\cdot p_c^{\rm in}-n\cdot p_s^{\rm tot})\,
   \delta^{(2)}(p_{c\perp}^{\rm out}-p_{c\perp}^{\rm in}
                -p_{s\perp}^{\rm tot}) \,.
\end{equation}
In each term we must eliminate one of the largest momentum components,
so that the remaining momenta are unconstrained. It follows that the 
first $\delta$-function eliminates an integral over a minus component of 
a collinear momentum, the second $\delta$-function eliminates an integral 
over a plus component of a soft momentum, and the last $\delta$-function
eliminates two components of a transverse momentum. It is important that
the second $\delta$-function must not be used to eliminate a plus 
component of a collinear momentum such as $n\cdot p_c^{\rm out}$, since 
for generic soft momenta the combination 
$n\cdot p_c^{\rm in}+n\cdot p_s^{\rm tot}$ would not be of order 
$E\lambda^2$. The eliminated momenta combined scale like $E^4\lambda^3$, 
and hence we must count the measure in the soft-collinear action as 
$\lambda^{-3}$. It follows that, at leading order in $\lambda$, only 
operators scaling like $\lambda^3$ can appear in the soft-collinear 
interaction Lagrangian.

\begin{figure}
\epsfxsize=11.0cm
\centerline{\epsffile{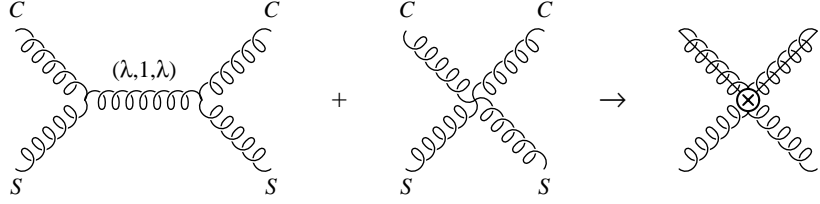}}
\vspace{0.0cm}
\centerline{\parbox{14cm}{\caption{\label{fig:softcol}
Effective four-gluon coupling obtained after integrating out the 
off-shell gluon in the first diagram and adding the corresponding local
QCD vertex. The resulting SCET operator is denoted by a crossed circle.
Collinear gluons in SCET are drawn as springs with a line. Soft and 
collinear gluons in full QCD are labeled by $S$ and $C$.}}}
\end{figure}

We start by discussing soft-collinear interactions induced by the 
exchange of off-shell modes with momentum scaling like 
$E(\lambda,1,\lambda)$, which are generically produced when soft fields
are coupled to collinear ones. In order to find the exact form of the 
corresponding terms in the Lagrangian one would have to integrate out 
the off-shell modes in the path integral. This is a difficult problem, 
whose solution we leave for future work. In the following we choose a 
``pedestrian'' approach and match the resulting interaction terms 
involving two soft and two collinear fields perturbatively. We find 
that at leading order in $\lambda$ there are interactions between two 
soft quarks and two collinear gluons, two collinear quarks and two soft 
gluons, and two soft and two collinear gluons. In the first two cases 
an off-shell quark propagator is integrated out, while in the latter 
case an off-shell gluon propagator occurs. As shown in 
Figure~\ref{fig:softcol}, in this case one must also include the local 
four-gluon vertex present in full QCD. We find that the resulting 
soft-collinear interactions connecting two collinear and two soft partons 
are obtained from the Lagrangian
\begin{eqnarray}\label{2gl}
   {\cal L}_{sc}^{({\rm induced})}
   &=& - g^2\,\bar q_s\,A_{c-}\,\frac{\nslash}{2}\,
    \frac{1}{i\partial_-}\,A_{c-}\,q_s
    - g^2\,\bar\xi\,A_{s+}\,\frac{\nbslash}{2}\,
    \frac{1}{i\partial_+}\,A_{s+}\,\xi \nonumber\\
   &-& \frac{g^2}{4}\,f_{abe} f_{mne}\,\Bigg\{
    A_{s+}^m\,\frac{1}{i\partial_+}\,A_{s+}^n \Big[
    A_{c-}^a\,i\partial_+\,A_{c-}^b
    - A_{c\perp\mu}^a\,i\partial_-\,A_{c\perp}^{\mu,b}
    + 2A_{c\perp\mu}^a\,i\partial_\perp^\mu\,A_{c-}^b \Big] \nonumber\\
   &&\hspace{1.835cm}\mbox{}+ A_{c-}^a\,\frac{1}{i\partial_-}\,A_{c-}^b
    \Big[ A_{s+}^m\,i\partial_-\,A_{s+}^n
    - A_{s\perp\mu}^m\,i\partial_+\,A_{s\perp}^{\mu,n}
    + 2A_{s\perp\mu}^m\,i\partial_\perp^\mu\,A_{s+}^n \Big] \nonumber\\
   &&\hspace{1.835cm}\mbox{}- \frac12\,i\partial_{\perp\mu} \Big(
    A_{s+}^m\,\frac{1}{i\partial_+}\,A_{s+}^n \Big)\,
    i\partial_\perp^\mu \Big( A_{c-}^a\,\frac{1}{i\partial_-}\,A_{c-}^b
    \Big) \Bigg\} \,,
\end{eqnarray}
where we use the short-hand notation $A_+=n\cdot A$ and 
$A_-=\bar n\cdot A$ etc.\ for brevity. Note that this Lagrangian is 
symmetric under the exchange of soft and collinear fields combined with 
$n\leftrightarrow\bar n$. That this is a symmetry of the soft-collinear 
interaction Lagrangian follows from the fact that a longitudinal Lorentz 
boost can be used to turn collinear fields into soft ones and vice versa.

\begin{figure}
\epsfxsize=3.5cm
\centerline{\epsffile{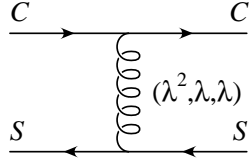}}
\vspace{0.0cm}
\centerline{\parbox{14cm}{\caption{\label{fig:messenger}
Example of a long-distance soft-collinear interaction induced by the 
exchange of a soft messenger gluon.}}}
\end{figure}

It is also possible to couple a single soft field to two or more 
collinear fields. In that case the $\delta$-functions in (\ref{delta}) 
enforce that the momentum of the soft field must scale like 
$E(\lambda^2,\lambda,\lambda)$. The smallness of the plus component of 
this momentum implies a phase-space suppression, which however is already
taken into account by assigning scaling $\lambda^{-3}$ rather than 
$\lambda^{-4}$ to the measure $d^4x$ in the action. Note that the soft 
parton is still off-shell by an amount of order $\Lambda^2$, as is any 
other soft mode. As illustrated in Figure~\ref{fig:messenger}, the soft 
field produced in such an interaction can interact with other soft 
particles, thereby acting as a messenger between the soft and collinear 
sectors of SCET. Let us add that, alternatively, we could study couplings 
of a single collinear field to two or more soft fields. In that case the 
$\delta$-functions in (\ref{delta}) enforce that the momentum of the 
collinear field must scale like $E(\lambda^2,\lambda,\lambda)$, and the 
phase-space suppression is now reflected in the smallness of the minus 
component of this momentum. In order to avoid double counting, let us 
agree that collinear fields always have a large momentum component, so a 
field with momentum scaling like $E(\lambda^2,\lambda,\lambda)$ is 
considered part of a soft mode, not a collinear one. This convention 
breaks the ``soft-collinear symmetry'' mentioned at the end of the last 
paragraph. It is nevertheless reasonable, since the $B$ meson defines a 
particular Lorentz frame, in which it is natural to consider a 
$(\lambda^2,\lambda,\lambda)$ mode as part of a soft mode. The asymmetry 
introduced by this choice could be avoided by introducing separate 
``soft-collinear'' fields for the $(\lambda^2,\lambda,\lambda)$ modes and 
studying their interactions with soft and collinear particles. But since 
we will show that these modes are irrelevant at leading order, this would 
only lead to an unnecessary proliferation of notation.

There are three elementary vertices in QCD which are of order $\lambda^3$ 
and couple a soft field to collinear fields: the coupling of a soft gluon 
to two collinear quarks, the coupling of a soft gluon to two collinear 
gluons, and the coupling of a soft gluon to three collinear gluons. These 
interactions follow from the Lagrangian
\begin{eqnarray}\label{1gl}
   {\cal L}_{sc}^{({\rm direct})}
   &=& g\,\bar\xi\,\frac{\nbslash}{2}\,A_{s+}\,\xi
    + \frac{g}{2}\,f_{abm}\,A_{s+}^m\,A_{c\mu}^a \big(
    2\partial^\mu A_{c-}^b - \partial_- A_c^{\mu,b} \big) \nonumber\\
   &-& \frac{g^2}{2}\,f_{abe} f_{mne}\,A_{c\mu}^a\,A_c^{\mu,m}\,
    A_{c-}^b\,A_{s+}^n \,.
\end{eqnarray}

An important observation following from the calculations presented above 
is that at leading order in $\lambda$ the soft-collinear interactions 
involve at least one $A_{s+}$ or $A_{c-}$ gluon field. We expect this 
feature to pertain also to interactions involving more than four partons. 
These interaction terms are ``unphysical'' in the sense that they involve 
scalar or longitudinal gluon polarizations and can be made to vanish by 
choosing the light-cone gauge conditions $n\cdot A_s=0$ and 
$\bar n\cdot A_c=0$. We will argue in Section~\ref{sec:pretty} that any 
gauge-invariant operator in SCET can be constructed out of 
gauge-invariant building blocks, which are non-zero in light-cone gauge.
This suggests that the interactions in (\ref{2gl}) and (\ref{1gl}) can be 
removed by field redefinitions (i.e., they vanish by the equations of 
motion). Let us demonstrate this explicitly for the terms involving quark 
fields. Consider the first term in (\ref{2gl}), which couples two 
collinear gluons to two soft quarks. Since we have derived this term by 
matching an amplitude with two external gluons, we are free to replace 
$i\partial_-$ in the denominator by a covariant derivative. Next we use 
the identity
\begin{equation}\label{niceid}
   g\bar n\cdot A_c\,q_s = - i\bar n\cdot D_c\,(W-1)\,q_s
   + (W-1)\,i\bar n\cdot\partial\,q_s \,,
\end{equation}
which follows from (\ref{neat}). The second term on the right-hand side 
is suppressed with respect to the first one by a power of $\lambda$ and 
can be neglected. Repeated application of this identity (and its Dirac 
conjugate) yields, to leading order in $\lambda$,
\begin{eqnarray}\label{interm}
   - g^2\,\bar q_s\,\frac{\nslash}{2}\,A_{c-}\,\frac{1}{i\partial_-}\,
    A_{c-}\,q_s 
   &\to& - \bar q_s\,\frac{\nslash}{2}\,(W^\dagger-1)\,
    i\bar n\cdot D_c\,(W-1)\,q_s \nonumber\\
   &=& - \bar q_s\,\frac{\nslash}{2}\,(W^\dagger\,i\bar n\cdot D_c\,W
    - i\bar n\cdot D_c)\,q_s
    - 2\,\bar q_s\,\frac{\nslash}{2}\,g\bar n\cdot A_c\,q_s \nonumber\\
   &=& - \bar q_s\,\frac{\nslash}{2}\,g\bar n\cdot A_c\,q_s \,.
\end{eqnarray}
This vanishes by momentum conservation (enforced by the integration over 
$d^4x$ in the action), since with our convention for collinear fields it
is impossible to couple a single collinear particle to soft fields. 
A derivation analogous to (\ref{interm}), but with all soft and collinear 
fields interchanged, can be used to show that the terms in (\ref{2gl}) 
and (\ref{1gl}) coupling two collinear quarks to one or two soft gluons
vanishes at leading order:
\begin{equation}
   - g^2\,\bar\xi\,A_{s+}\,\frac{\nbslash}{2}\,
   \frac{1}{i\partial_+}\,A_{s+}\,\xi
   + g\,\bar\xi\,\frac{\nbslash}{2}\,A_{s+}\,\xi \to 0 \,.
\end{equation}
Similar arguments should apply for the pure-glue interactions. We 
conclude that the soft-collinear Lagrangian in (\ref{Leff}) vanishes to 
leading order, ${\cal L}_{sc}^{({\rm LO})}=0$, and that the 
$(\lambda^2,\lambda,\lambda)$ messenger modes of the type shown in 
Figure~\ref{fig:messenger} are irrelevant to leading power in $\lambda$. 
These findings agree with arguments based on soft-collinear gauge 
invariance presented in \cite{Bauer:2001yt}, where however the special 
kinematics of soft-collinear interactions and the potential relevance of 
messenger modes were not addressed. The observation that soft-collinear 
interactions are absent at leading power in the SCET Lagrangian will be 
of crucial importance to the discussion of factorization in 
Sections~\ref{sec:4qops} and~\ref{sec:examples}. 

We stress at this point that non-trivial soft-collinear interactions do
occur at next-to-leading order in $\lambda$. For instance, a 
gauge-invariant coupling of two soft quarks to two collinear gluons can 
be written in the form
\begin{equation}
   - \bar q_s\,S\,\calAslash_{c\perp}\,\frac{\nslash}{2}\,
   \frac{1}{in\cdot\partial}\,\calAslash_{c\perp}\,S^\dagger q_s \,,
\end{equation}
which cannot be made to vanish in any gauge. A similar interaction (with
soft and collinear fields interchanged) can also be written for the 
coupling of two collinear quarks to two soft gluons. These terms are of 
order $\lambda^4$ and contribute at subleading power to the Lagrangian 
${\cal L}_{sc}$. Because the soft and collinear Lagrangians (\ref{Lsoft}) 
and (\ref{Lcol}) are exact, and the heavy-quark Lagrangian (\ref{HQET}) 
is known beyond the leading order, a complete derivation of the 
soft-collinear interaction terms of order $\lambda^4$ is the last missing 
step in the construction of the SCET Lagrangian at next-to-leading order. 
We leave this derivation for future work.

\section{Soft-collinear currents}
\label{sec:softcol}

An important application of the SCET formalism concerns the 
representation of external current operators (such as the flavor-changing 
operators arising in weak interactions) in terms of effective-theory 
fields. For the case of heavy-light currents the corresponding matching 
relation is \cite{Bauer:2001yt}
\begin{equation}\label{HLmatch}
   \big[ \bar\psi(x)\,\Gamma\,b(x) \big]_{\rm QCD} 
   \to e^{-i m_b v\cdot x} \sum_i C_i(m_b,E,\mu)\,
   \big[ \bar\xi\,W \big](x)\,\Gamma_i\,\big[ S^\dagger h \big](x)
   + \dots \,,
\end{equation}
where $E=\bar n\cdot p_c^{\rm tot}$ is the large component of the total 
collinear momentum, which is fixed by kinematics, and the dots represent 
higher-order terms in $\lambda$. The Dirac matrices $\Gamma_i$ have the 
same transformation properties as the original matrix $\Gamma$. For 
instance, in the case of a vector current we have $\Gamma=\gamma^\mu$ and 
$\Gamma_i=\{\gamma^\mu,v^\mu,n^\mu\}$. The SCET current operators are 
manifestly invariant under soft and collinear gauge transformations. Note 
that the string operators $W$ and $S$ sum up an infinite set of couplings 
(involving collinear and soft gluons) allowed at leading power. That this 
is done correctly can be checked explicitly by perturbative matching 
\cite{Bauer:2000yr}. When a soft gluon couples to a collinear quark it 
produces a fluctuation with momentum scaling like $E(\lambda,1,\lambda)$, 
which is off-shell by an amount of order $E\Lambda$. Similarly, when a 
collinear gluon couples to a heavy quark it produces a fluctuation that 
is off-shell by an amount of order $E^2$. These modes remain off their 
mass shell when further soft or collinear gluons are coupled to them. 
When the off-shell modes are integrated out in the path integral, one 
reproduces the product of the two Wilson lines in (\ref{HLmatch}) 
\cite{Bauer:2001yt}. It is sufficient to keep the leading terms in the 
$\lambda$ expansion at any stage in this derivation. The propagator for 
an off-shell collinear quark scales like $1/\Lambda$, and it combines 
with the soft gluon field $n\cdot A_s\sim\Lambda$ to give a leading-order 
contribution (other components of the soft gluon field do not contribute 
at leading power, since 
$\bar\xi\,\Aslash_s\,\nslash=2\,\bar\xi\,n\cdot A_s$). Similarly, the 
propagator for an off-shell heavy quark scales like $1/E$, and so only 
the leading component $\bar n\cdot A_c\sim E$ of the collinear gluon 
field must be kept. 

Let us now study the analogous situation in which the soft heavy quark is
replaced by a soft light quark. (We have been unable to find a 
phenomenological application of the resulting light-light soft-collinear 
current, so that this example is of academic value. Nevertheless, it will 
elucidate the novel features encountered in the presence of soft light 
quarks. The results derived in this section can readily be generalized to 
more realistic situations.) The naive guess 
$\bar\xi\,W\,\Gamma_i\,S^\dagger q_s$ for the resulting current operators 
in SCET is wrong for two reasons: first, the presence of an intermediate 
mass scale leads to a non-locality of the resulting operators at 
large scales of order $1/\Lambda$; secondly, a more complicated structure 
of collinear fields is induced in the matching process.

The appearance of an intermediate scale can already be seen at one-loop 
order in perturbation theory. When the matching is performed using 
on-shell external quark states (which is legitimate, since the Wilson 
coefficients are insensitive to infrared physics) with momenta $l$ (soft) 
and $p_c=E n$ (collinear), the only non-zero invariant is 
$l\cdot p_c=E\,n\cdot l$, which is a perturbative scale of order 
$E\Lambda$. However, the value of this scale depends on non-perturbative 
hadronic physics through its dependence on a component of a soft 
momentum. A one-loop matching calculation yields an expression of the 
form (the Dirac structure is preserved, so there is no need for a 
summation over matrices $\Gamma_i$)
\begin{equation}
   C_\Gamma\bigg(\frac{E l_+}{\mu^2}\bigg)\,\bar u_\xi(p_c)\,
  \Gamma\,u_{q_s}(l) \,,
\end{equation}
where $l_+=n\cdot l-i0$. In order to write this as the matrix element of 
an operator we replace $l_+$ by a derivative on the light-quark field and 
obtain
\begin{equation}
   \int dl_+\,C_\Gamma\bigg(\frac{E l_+}{\mu^2}\bigg)\,
   \bar\xi\,\Gamma\,\delta(l_+-in\cdot\partial)\,q_s \,.
\end{equation}
The $\delta$-function indicates that the resulting operator is non-local 
on a scale of order $1/\Lambda$. Introducing the Fourier transform of the 
Wilson coefficient, the above result can be rewritten as
\begin{equation}\label{LLmatch1}
   \int dt\,\widetilde C_\Gamma(t,E,\mu)\,
   \big[ \bar\xi\,W \big](x)\,\Gamma\,\big[ S^\dagger q_s \big](x+tn) \,,
\end{equation}
where
\begin{equation}
   \widetilde C_\Gamma(t,E,\mu)
   = \frac{1}{2\pi} \int dl_+\,e^{il_+ t}\,
   C_\Gamma\bigg(\frac{E l_+}{\mu^2}\bigg) \,.
\end{equation}
The string operators $W$ and $S$ have been inserted here so that the 
resulting expression is gauge invariant. Note that by definition $l_+$ is 
now the plus component of the total momentum carried by the soft fields
$S^\dagger q_s$. Likewise, $2E$ is the minus component of the total 
momentum carried by the collinear fields $\bar\xi\,W$. The non-locality 
of the soft-collinear current operator in SCET is a novel feature, which 
makes (\ref{LLmatch1}) more complicated than the corresponding expression 
(\ref{HLmatch}) for heavy-light currents. (At tree level, however, 
$C_\Gamma=1$ and $\widetilde C_\Gamma=\delta(t)$, so the non-locality is 
absent.) In the discussion of four-quark operators in 
Section~\ref{sec:4qops}, this will naturally introduce gauge-invariant 
soft quark bilinears of the form 
$[\bar h\,S](0)\dots [S^\dagger q_s](tn)$. The $B$-meson matrix elements 
of such operators define the leading-order light-cone distribution 
amplitudes \cite{Grozin:1996pq}.

\begin{figure}
\epsfxsize=3.5cm
\centerline{\epsffile{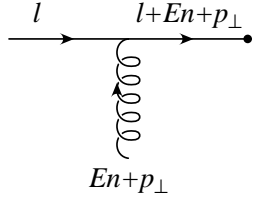}}
\vspace{0.0cm}
\centerline{\parbox{14cm}{\caption{\label{fig:vertex1}
Attachment of a collinear gluon to a soft light quark.}}}
\end{figure}

\begin{figure}
\epsfxsize=12.0cm
\centerline{\epsffile{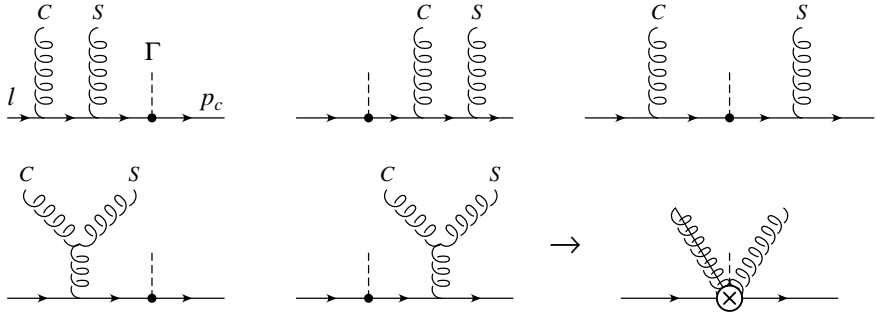}}
\vspace{0.0cm}
\centerline{\parbox{14cm}{\caption{\label{fig:2gluon}
Diagrams contributing to the matching calculation for the soft-collinear 
current for the case of one soft and one collinear external gluon, and 
the resulting non-local operator in the SCET. The dashed line represents 
the current insertion.}}}
\end{figure}

Surprisingly, eq.~(\ref{LLmatch1}) is still not the final answer for the 
representation of the current in SCET. In order to understand this, let 
us study in more detail what happens when a collinear gluon hits a 
massless soft quark. As indicated in Figure~\ref{fig:vertex1}, the 
resulting off-shell mode has momentum $l+En+p_\perp$ scaling like 
$E(\lambda,1,\lambda)$. Keeping terms up to subleading order in 
$\lambda$, the corresponding propagator and vertex yield
\begin{equation}\label{graphs}
   \frac{i(E\nslash+\pslash_\perp+\lslash)}{2En\cdot l}\,ig
    \left( \frac{\nslash}{2}\,\bar n\cdot A_c + \Aslash_{c\perp}
    \right) q_s
   = -g \left( \frac{\bar n\cdot A_c}{2E} + \frac{\nslash}{2l_+}\,
    \bigg[ \Aslash_{c\perp} - \pslash_\perp\,\frac{\bar n\cdot A_c}{2E}
    \bigg] \right) q_s \,,
\end{equation}
where we have used the equation of motion $\lslash\,q_s=0$ for the light 
quark. The key point to note about this result is that the off-shell
propagator scales like $1/\Lambda$, while the largest component of the 
collinear gluon field scales like $E$. The superficially largest term of
order $E/\Lambda$ (which would upset power counting) cancels. However, 
leading contributions arise from the subleading terms in both the 
propagator and the gluon field. While the first term in the above result 
corresponds to the expansion of the Wilson line $W$, the remaining terms 
correspond to the object $(i\Dslash_{c\perp} W)$. The factor $1/l_+$ 
associated with these terms gives rise to a non-locality even at tree 
level. A careful analysis (using perturbative matching) reveals that 
there is a second type of current operator in the effective theory, given 
by
\begin{equation}\label{LLmatch2}
   -\int dt\,ds\,\widetilde D_\Gamma(t,s,E,\mu)\,
   \big[ \bar\xi\,W \big](x)\,\Gamma\,\frac{\nslash}{2}\,
   \calAslash_{c\perp}(x+s\bar n)\,\big[ S^\dagger q_s \big](x+tn) \,,
\end{equation}
where $\A_c^\mu$ is the gauge-invariant object defined in (\ref{Adef}). 
The Wilson coefficient $\widetilde D_\Gamma$ is related by a double 
Fourier transformation to a momentum-space coefficient function,
\begin{equation}
   \widetilde D_\Gamma(t,s,E,\mu) 
   = \frac{1}{(2\pi)^2} \int dl_+\,dp_-\,e^{il_+ t}\,e^{-ip_- s}\,
   D_\Gamma(l_+,p_-,E,\mu) \,,
\end{equation}
where $p_-=\bar n\cdot p_g$ is the minus component of the collinear 
momentum carried by the field $\A_c$, and 
$E=\frac12\,\bar n\cdot p_c^{\rm tot}$ is the large energy carried by all
collinear fields. At tree level, we find $D_\Gamma=1/l_+$ and hence
$\widetilde D_\Gamma=i\theta(t)\,\delta(s)$. We have checked by explicit 
calculation that the sum of the two expressions in (\ref{LLmatch1}) and 
(\ref{LLmatch2}) reproduces (at tree level) the current matrix elements 
with an arbitrary number of collinear gluons and no soft gluon, an 
arbitrary number of soft gluons and no collinear gluon, and one soft and 
one collinear gluon. The relevant diagrams for the latter case are shown 
in Figure~\ref{fig:2gluon}. One must add to these graphs a contribution 
from the equation of motion $i\delslash\,q_s=-g\Aslash_s\,q_s$ for the 
soft quark applied to the graph with only one external collinear gluon. 
Note that the transverse collinear gluon field in (\ref{LLmatch2}) 
appears together with a factor of $\nslash$. We will show in the next 
section that this is, in fact, required by reparameterization invariance. 
It is therefore not possible to obtain more than one insertion of the 
product $\nslash\,\calAslash_{c\perp}$. Eq.~(\ref{LLmatch2}) then gives 
the most general operator with a transverse collinear gluon insertion 
that is allowed by gauge and reparameterization invariance.

\begin{figure}
\epsfxsize=8.0cm
\centerline{\epsffile{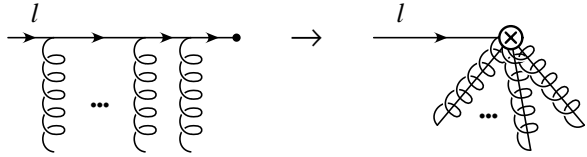}}
\vspace{0.0cm}
\centerline{\parbox{14cm}{\caption{\label{fig:vertex}
Multiple attachments of collinear gluons to a soft light quark, and the 
resulting non-local interaction in SCET.}}}
\end{figure}

A formal argument justifying the result (\ref{LLmatch2}) for an arbitrary 
number of soft and collinear gluons would have to be based on integrating 
out the off-shell $(\lambda,1,\lambda)$ modes of quarks and gluons in the
path integral, generalizing the discussion presented in Appendix~A of 
\cite{Bauer:2001yt}. This is more complicated in the present case, 
however, because the transverse components and transverse derivatives of 
the gauge fields cannot be ignored. For the sake of simplicity we 
consider here only the case of an arbitrary number of collinear gluons 
attached to a soft quark, ignoring soft gluons. We then need to integrate 
out off-shell $(\lambda,1,\lambda)$ modes $\psi_X$ of the light quark, 
but there is no need to consider off-shell gluon fields. The QCD 
Lagrangian gives rise to the following interactions between the relevant 
on-shell and off-shell fields:
\begin{equation}
   {\cal L}_X = \bar\psi_X\,i\Dslash_c\,\psi_X
   + \bar\psi_X\,g\Aslash_c\,q_s + \bar q_s\,g\Aslash_c\,\psi_X \,.
\end{equation}
In the kinetic term for the off-shell field the full covariant derivative 
should appear, but within the approximation just described we need to 
keep only the collinear gauge field. Since the off-shell field carries a 
large momentum in the $n$ direction it is useful to split it up into two 
2-component fields $\psi_n$ and $\psi_{\bar n}$ defined such that 
$\nslash\,\psi_n=0$ and $\nbslash\,\psi_{\bar n}=0$. The equations of 
motion satisfied by these fields are
\begin{equation}\label{EoM}
\begin{aligned}
   in\cdot D_c\,\psi_n
   + \frac{\nslash}{2}\,i\Dslash_{c\perp}\,\psi_{\bar n}
   + \frac{\nslash}{2} \left( g\Aslash_{c\perp}
   + \frac{\nbslash}{2}\,gn\cdot A_c \right) q_s &= 0 \,, \\
   i\bar n\cdot D_c\,\psi_{\bar n}
   + \frac{\nbslash}{2}\,i\Dslash_{c\perp}\,\psi_n
   + \frac{\nbslash}{2} \left( g\Aslash_{c\perp}
   + \frac{\nslash}{2}\,g\bar n\cdot A_c \right) q_s &= 0 \,.
\end{aligned}
\end{equation}
Solving the equation for $\psi_{\bar n}$ to leading order in $\lambda$ we 
obtain
\begin{equation}\label{sol1}
   \psi_{\bar n} = - \frac{\nbslash\,\nslash}{4}\,
   \frac{1}{i\bar n\cdot D_c}\,g\bar n\cdot A_c\,q_s + \dots
   = \frac{\nbslash\,\nslash}{4}\,(W-1)\,q_s + \dots \,, \\
\end{equation}
where the dots denote higher-order terms, and to arrive at the second 
equality we have used the identity (\ref{niceid}). Inserting the solution 
for $\psi_{\bar n}$ into the first equation in (\ref{EoM}) yields
\begin{eqnarray}\label{sol2}
   in\cdot D_c\,\psi_n
   &=& - \frac{\nslash}{2}\,(i\Dslash_{c\perp} W)\,q_s
    - \frac{\nslash}{2}\,(W-1)\,i\delslash_\perp\,q_s + \dots \nonumber\\
   &=& - \frac{\nslash}{2}\,(i\Dslash_{c\perp} W)\,q_s
    + \frac{\nslash\,\nbslash}{4}\,(W-1)\,in\cdot\partial\,q_s
    + \dots \,,
\end{eqnarray}
where in the last step the equation of motion $i\delslash\,q_s=0$ for the 
light quark (in the absence of soft gluons) has been used. Because the 
off-shell field has momentum $n\cdot p_X\sim\lambda$, which is larger 
than $n\cdot A_c\sim\lambda^2$, we can replace the covariant derivative 
on the left-hand side of this equation by an ordinary derivative. 
Combining then the two results in (\ref{sol1}) and (\ref{sol2}), and 
using that the derivative $in\cdot\partial$ commutes with collinear 
fields to leading order in $\lambda$, we obtain
\begin{equation}\label{psiX}
   \psi_X = (W-1)\,q_s - \frac{\nslash}{2}\,(i\Dslash_{c\perp} W)\,
   \frac{1}{in\cdot\partial}\,q_s + \dots \,.
\end{equation}
Figure~\ref{fig:vertex} illustrates that this result can be understood as 
a tree-level matching relation for a light quark (in the absence of soft 
gluons),
\begin{eqnarray}\label{qmatch}
   q(x)\big|_{\rm QCD}\to \big[ q_s + \psi_X \big](x)
   &=& W \left( 1 - \frac{\nslash}{2}\,\big[
    W^\dagger(i\Dslash_{c\perp} W) \big]\,
    \frac{1}{in\cdot\partial-i\epsilon} \right) q_s \nonumber \\
   &=& W(x) \left[ q_s(x) - i\,\frac{\nslash}{2}\,
    \calAslash_{c\perp}(x) \int_0^\infty\!dt\,q_s(x+tn) \right] ,
\end{eqnarray}
where we have chosen a $-i\epsilon$ prescription to regularize the 
inverse derivative on the light-quark field. This choice is consistent 
with the Feynman prescription for the propagator of an off-shell quark
obtained by coupling a final-state collinear gluon to an initial state 
soft quark. Performing this replacement, along with $\psi\to\xi$, in the 
QCD current operator $\bar\psi\,\Gamma\,q$ precisely reproduces the 
tree-level structure of collinear fields in (\ref{LLmatch1}) and 
(\ref{LLmatch2}). The extension of this argument to include soft gluon 
fields is left for future work. Note that the integro-differential 
operator on the right-hand side of the first equation in (\ref{qmatch}) 
is nilpotent in the sense that $\exp(1-\dots)=(1-\dots)$. Therefore, 
insertions of transverse collinear gluons on a soft light-quark line do 
not exponentiate.

It is, at first sight, surprising that the operator in (\ref{LLmatch2}) 
is of leading order in SCET power counting, because the extra transverse 
gluon field $\calAslash_{c\perp}$, which is absent in (\ref{LLmatch1}), 
scales like a power of $\lambda$. However, this suppression is 
compensated by the different behavior of the Wilson coefficient 
functions, $\widetilde C_\Gamma\sim\lambda$ and 
$\widetilde D_\Gamma\sim 1$. (In momentum space, the transverse 
derivative is compensated by the factor $1/l_+$ associated with the 
transverse terms in (\ref{graphs}).) It is evident from this example that 
the presence of large non-localities on the scale $1/\Lambda$ can upset 
naive SCET power counting. From our discussion so far it follows that 
one needs to know the $t$-dependence of the short-distance coefficients 
(or the $l_+$ dependence of the corresponding coefficients in momentum 
space) before one can decide whether a non-local operator such as 
(\ref{LLmatch1}) or (\ref{LLmatch2}) contributes at leading order in 
power counting. Fortunately, it is possible to deduce the $t$-dependence 
of the Wilson coefficients to all orders in perturbation theory without 
an explicit calculation. This is discussed in the next section.

\section{Reparameterization invariance}
\label{sec:RPI}

Operators in SCET must be invariant under redefinitions of the light-cone 
basis vectors $n$ and $\bar n$ that leave the scaling properties of 
fields and momenta unchanged \cite{Chay:2002vy,Manohar:2002fd}. This 
property is referred to as reparameterization invariance, and it can be 
used to derive constraints on the Wilson coefficients of SCET operators, 
often relating the coefficients of some operators to those of others. 
Reparameterization invariance is a consequence of the invariance of QCD 
under Lorentz transformations, which is not explicit (but still present) 
in SCET because of the introduction of the light-cone vectors.

It is useful to distinguish between three classes of infinitesimal 
transformations, corresponding to two different transverse boosts and a 
longitudinal boost:
\begin{equation}
\begin{aligned}
   &\mbox{Type~I:} & \qquad &n^\mu\to n^\mu + \epsilon_\perp^\mu \,,
   & \qquad &\bar n^\mu \mbox{~invariant} & \qquad
    &\mbox{(with $\epsilon_\perp^\mu\sim\lambda$)} \\
   &\mbox{Type~II:} & &\bar n^\mu\to \bar n^\mu + e_\perp^\mu \,, &
    &n^\mu \mbox{~invariant} &
    &\mbox{(with $e_\perp^\mu\sim 1$)} \\
   &\mbox{Type~III:} & &n^\mu\to n^\mu/\alpha \,, &
    &\bar n^\mu\to \alpha\bar n^\mu &
    &\mbox{(with $\alpha\sim 1$)}
\end{aligned}
\end{equation}
In parenthesis we give the scaling properties for the parameters of the 
corresponding finite transformations, which are relevant for power 
counting. Using the properties of the fields and Wilson lines under these 
transformation, as compiled in Table~I of \cite{Manohar:2002fd}, it is 
straightforward to show that the current operators in (\ref{LLmatch1}) 
and (\ref{LLmatch2}) are separately invariant under type~I and type~II 
transformations to leading order in $\lambda$. In other words, 
reparameterization invariance links these operators with operators that 
appear at subleading order in $\lambda$. The only non-trivial point in 
this analysis concerns the transverse collinear derivative 
$D_{c\perp}^\mu$, which has non-vanishing variations at leading order in 
$\lambda$ under both type~I and type~II transformations,
\begin{equation}
   D_{c\perp}^\mu \stackrel{{\rm type~I}}{\to} D_{c\perp}^\mu
   - \frac{\epsilon_\perp^\mu}{2}\,\bar n\cdot D_c + O(\lambda^2) \,,
   \qquad
   D_{c\perp}^\mu \stackrel{{\rm type~II}}{\to} D_{c\perp}^\mu
   - \frac{n^\mu}{2}\,e_\perp\cdot D_{c\perp} + O(\lambda^2) \,.
\end{equation} 
However, in both cases the object 
$\nslash\,W^\dagger(i\Dslash_{c\perp} W)=\nslash\,\calAslash_{c\perp}$ is 
left invariant. For type~I transformations this follows from 
$(\bar n\cdot D_c\,W)=0$, whereas for type~II transformations it follows 
since $\nslash^2=0$. We stress that, without the extra factor of 
$\nslash$, the operator in (\ref{LLmatch2}) would not be invariant under 
a type~II reparameterization.

The type~III transformations have non-trivial consequences. We find that 
the current operators are invariant under these transformations only if 
their Wilson coefficients obey the homogeneity relations
\begin{equation}
   \widetilde C_\Gamma(t,E,\mu)
   = \alpha\,\widetilde C_\Gamma(\alpha t,\alpha E,\mu) \,, \qquad
   \widetilde D_\Gamma(t,s,E,\mu) = \frac{1}{\alpha}\,
    \widetilde D_\Gamma(\alpha t,s/\alpha,\alpha E,\mu) \,.
\end{equation}
Taking into account the canonical dimensions of these coefficients, it 
follows that
\begin{equation}\label{RPIpred}
\begin{aligned}
   \widetilde C_\Gamma(t,E,\mu)
   &= \delta(t)\,c_\Gamma^{(1)}[\alpha_s(\mu)] + \frac{1}{t}\,
    c_\Gamma^{(2)}\big[\mu^2\,t/E,\alpha_s(\mu)\big] \,, \\
   \widetilde D_\Gamma(t,s,E,\mu)
   &= \delta(s)\,d_\Gamma^{(1)}\big[\mu^2\,t/E,\alpha_s(\mu)\big] 
    + \frac{1}{s}\,
    d_\Gamma^{(2)}\big[\mu^2\,t/E,sE,\alpha_s(\mu)\big] \,,
\end{aligned}
\end{equation}
where the coefficient functions $c_\Gamma^{(i)}$ and $d_\Gamma^{(i)}$ are 
dimensionless. Since the dependence of the Wilson coefficients on the 
renormalization scale is logarithmic, we conclude that to all orders of 
perturbation theory $\widetilde C_\Gamma(t,E,\mu)\sim 1/t\sim\Lambda$ and 
$\widetilde D_\Gamma(t,s,E,\mu)\,ds\sim 1$ modulo logarithms. With this 
information, it is now evident that the two types of current operators 
contribute at the same order in power counting.

The above argument based on longitudinal boost invariance determines the 
behavior of the momentum-space coefficients on the soft momentum $l_+$ to 
all orders in perturbation theory. This provides valuable information 
about the convergence of convolution integrals of hard-scattering kernels 
(Wilson coefficients) with the $B$-meson light-cone distribution 
amplitudes, which will be an important ingredient to factorization 
proofs.

\begin{figure}
\epsfxsize=16.0cm
\centerline{\epsffile{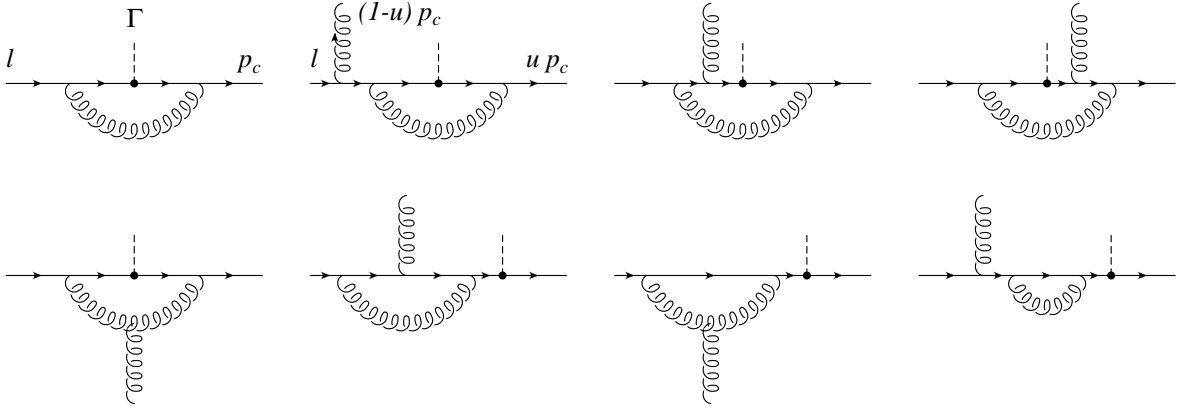}}
\vspace{0.0cm}
\centerline{\parbox{14cm}{\caption{\label{fig:current}
One-loop diagrams required to determine the matching coefficients
$C_\Gamma$ and $D_\Gamma$. The first diagram suffices to find the 
coefficient $C_\Gamma$, while the evaluation of the remaining graphs with 
an external collinear gluon is required to obtain $D_\Gamma$.}}}
\end{figure}

It may be instructive to illustrate our results for the soft-collinear 
current with a concrete example. For the case of the vector current, the 
explicit expressions for the matching coefficients $C_V$ and $D_V$ 
obtained at one-loop order (in the $\overline{\mbox{MS}}$ scheme, but
before subtraction of the pole terms in $\epsilon=2-d/2$) read 
\begin{equation}\label{coefs}
\begin{aligned}
   C_V &= 1 + \frac{C_F\,\alpha_s(\mu)}{4\pi}
   \left( \frac{2El_+}{\mu^2} \right)^{-\epsilon}
   \left( - \frac{2}{\epsilon^2} - \frac{3}{\epsilon} - 8
   + \frac{\pi^2}{6} \right) , \\
   D_V &= \frac{1}{l_+}\,\bigg\{ 1 + \frac{\alpha_s(\mu)}{4\pi}
    \left( \frac{2El_+}{\mu^2} \right)^{-\epsilon}
    \left[ C_F\,k_F(u,\epsilon) - \frac{C_A}{2}\,k_A(u,\epsilon)
    \right] \bigg\} \,,
\end{aligned}
\end{equation}
where
\begin{eqnarray}
   k_F(u,\epsilon) &=& - \frac{2}{\epsilon^2}
    - \frac{1}{\epsilon} \left( 1 - \frac{2\ln u}{1-u} \right)
    - \frac{\ln^2 u}{1-u} + \frac{4\ln u}{1-u} - 3 + \frac{\pi^2}{6} \,,
    \nonumber\\
   k_A(u,\epsilon) &=& \frac{1}{\epsilon}\,\frac{2\ln u}{1-u}
    - \frac{\ln^2 u}{1-u} + \frac{4\ln u}{1-u} + \frac{2\ln(1-u)}{u} \,.
\end{eqnarray}
The coefficient $D_V$ depends in a non-trivial way on the fraction $u$ of 
the total collinear momentum carried by the collinear quark (we define
$\bar n\cdot p_q=2uE$ and $\bar n\cdot p_g=2(1-u)E$). To obtain these 
results we work with on-shell external quark and gluon states and use 
dimensional regularization to regulate both ultraviolet and infrared 
divergences. This ensures that all SCET loop diagrams vanish (since 
there is no large scale left in the low-energy theory), and so the 
matching calculation is reduced to the calculation of vertex graphs in 
the full theory \cite{Eichten:1990vp,Neubert:1993za}. While the 
computation of $C_V$ is a simple exercise, to obtain $D_V$ requires 
evaluating the vertex and box diagrams with an external collinear gluon 
shown in Figure~\ref{fig:current}. A subtle point in this calculation is
that there is a contribution to $D_V$ resulting from the application of 
the equation of motion for the collinear quark in the first diagram. It 
is a highly non-trivial check of our result that the sum of the seven 
diagrams with an external gluon can be represented as the sum of two 
contributions corresponding to the two operators in (\ref{LLmatch1}) and 
(\ref{LLmatch2}), with the coefficient $C_V$ of the first operator fixed 
by the $u$-independent expression in (\ref{coefs}) obtained from the 
first diagram shown in the figure.

Taking the Fourier transforms of these results, we find the 
position-space coefficient functions
\begin{equation}
\begin{aligned}
   \widetilde C_V &= \delta(t) + \frac{C_F\,\alpha_s(\mu)}{4\pi}\,
   \frac{\theta(t)}{t}
   \left( \frac{\mu^2\,it\,e^{\gamma_E}}{2E}
   \right)^\epsilon \epsilon \left( -\frac{2}{\epsilon^2}
   - \frac{3}{\epsilon} -8 + \frac{\pi^2}{6} \right) , \\
  \widetilde D_V &= i\theta(t)\,\Bigg\{ \delta(s)
   + \frac{\alpha_s(\mu)}{4\pi}\,
   \left( \frac{\mu^2\,it\,e^{\gamma_E}}{2E} \right)^\epsilon \\[-0.2cm]
  &\hspace{2.47cm}\times
   \frac{E}{\pi} \int du\,e^{-2iEs(1-u)}\,\left[ 
   C_F\,k_F(u,\epsilon) - \frac{C_A}{2}\,k_A(u,\epsilon) \right]
   \Bigg\} \,,
\end{aligned}
\end{equation}
in accordance with the general forms in (\ref{RPIpred}) predicted by 
reparameterization invariance. When taking the limit $\epsilon\to 0$ in
the first result the pole at $t=0$ must be regularized, e.g.\ by using a 
plus distribution. In general, this will lead to an extra $1/\epsilon$
pole, which cancels the factor of $\epsilon$ encountered in the process
of Fourier transformation.

The scale dependence of the Wilson coefficient functions (in momentum or
position space) is governed by complicated, integro-differential 
renormalization-group equations \cite{Grozin:1996pq}. The ``anomalous 
dimension kernels'' in these equations contain a logarithm of the ratio 
of $\mu^2$ to the intermediate scale $E l_+$. The integration of the 
renormalization-group equations leads to the exponentiation of the 
leading Sudakov double logarithms (see, e.g., 
\cite{Bauer:2000ew,Bauer:2000yr}).

\section{Matching of four-quark operators}
\label{sec:4qops}

The discussion of the soft-collinear current presented in the previous 
two sections has elucidated many new features encountered in the 
interactions of a soft light quark with collinear gluons. We will now
consider a slightly more complicated example of a matching calculation,
which however is of greater phenomenological importance. Our goal is to 
match a local color-singlet four-quark operator of the type 
$O_{4q}=\bar\psi\,\Gamma_1\,T_1\,\psi\,\bar b\,\Gamma_2\,T_2\,q$ onto 
operators in SCET, in a kinematical situation where the quarks $q$ and 
$\bar b$ are soft, while $\psi$ and $\bar\psi$ are collinear. Here 
$\Gamma_{1,2}$ are arbitrary Dirac structures, and 
$T_1\otimes T_2={\bf 1}\otimes{\bf 1}$ or $t_a\otimes t_a$ are color 
structures. The resulting operators in SCET have the spinor content 
$\bar\xi\dots\xi\,\bar h\dots q_s$. Such operators would arise, e.g., in 
the discussion of factorization for the hard-scattering term in the 
exclusive decay $B\to K^*\gamma$. (However, in general they are obtained
by matching a non-local full-theory amplitude onto a four-quark operator
in the SCET.)

The matching calculation for four-quark operators proceeds in analogy to 
the matching for the soft-collinear current discussed in 
Section~\ref{sec:softcol}. We find that again two types of operators 
appear at leading order in $\lambda$. At tree level, the result is 
(setting $x=0$ for simplicity)
\begin{eqnarray}
   O_{4q}(0) &\to& \big[ \bar\xi\,W\,\Gamma_1\,T_1\,W^\dagger\xi
    \big](0)\,\big[ \bar h\,S\,\Gamma_2\,T_2\,S^\dagger q_s \big](0)
    \nonumber\\
   &&\mbox{}- \frac{i}{2} \int_0^\infty\!dt\,
    \big[ \bar\xi\,W\,\Gamma_1\,T_1\,W^\dagger\xi \big](0)\,
    \big[ \bar h\,S\,\Gamma_2\,T_2\,\nslash\,\calAslash_{c\perp}
    \big](0)\,\big[ S^\dagger q_s \big](tn) \,.
\end{eqnarray}
For final states containing up to two external gluons, we have checked 
the correctness of this expression explicitly using perturbative 
matching. Note, in particular, that the structure of collinear gluon 
fields follows from the matching relation (\ref{qmatch}).

When radiative corrections are included, the above result gets 
generalized in several ways. First, Dirac structures different from those 
of the original operator can be induced. Secondly, both color structures 
arise, since they mix under renormalization. Finally, the various 
components of the SCET operators become non-local. The most general
gauge-invariant matching relation at leading order in $\lambda$ is of the 
form
\begin{eqnarray}\label{4qfinal}
  O_{4q}(0) 
  &\to& \sum_{i,j} \sum_{C=S,O} \Bigg\{ \int dr\,dt\,
   \widetilde C_{ij}^{(C)}(r,t,E,m_b,\mu)\,Q_{ij}^{(C)}(r,t) \nonumber\\
  &&\hspace{1.28cm}\mbox{}- \frac12 \int dr\,ds\,dt\,
   \widetilde D_{ij}^{(C)}(r,s,t,E,m_b,\mu)\,R_{ij}^{(C)}(r,s,t) \Bigg\} 
   \,,
\end{eqnarray}
where
\begin{equation}\label{QRdef}
\begin{aligned}
   Q_{ij}^{(C)}(r,t) &= \big[ \bar\xi\,W \big](-r\bar n)\,\Gamma_i\,T_1\,
    \big[ W^\dagger\xi \big](r\bar n)\,
    \big[ \bar h\,S \big](0)\,\Gamma_j\,T_2\,
    \big[ S^\dagger q_s \big](tn) \,, \\
   R_{ij}^{(C)}(r,s,t) &= \big[ \bar\xi\,W \big](-r\bar n)\,\Gamma_i\,
    T_1\,\big[ W^\dagger\xi \big](r\bar n)\,
    \big[ \bar h\,S \big](0)\,\Gamma_j\,T_2\,
    \nslash\,\calAslash_{c\perp}(s\bar n)\,\big[ S^\dagger q_s \big](tn)
\end{aligned}
\end{equation}
are non-local operators, and the color label $C=S$ or $O$ refers to the 
color singlet-singlet and color octet-octet structures, respectively. As
in the case of the soft-collinear current both operators contribute at 
the same order in $\lambda$, since $\widetilde C_{ij}^{(C)}\sim\lambda$ 
while $\widetilde D_{ij}^{(C)}\sim 1$.

The form of the operators in (\ref{4qfinal}) is determined by gauge 
invariance. To see this, note that instead of working in the 
singlet--octet basis of operators with flavor structure 
$\bar\xi\dots\xi\,\bar h\dots q_s$ we could have chosen instead to work 
with operators containing only products of color-singlet currents. Gauge 
invariance would then require that these operators be of the type
\begin{equation}
\begin{aligned}
   &\big[ \bar\xi\,W \big]\,\Gamma_i\,\big[ W^\dagger\xi \big]\,
    \big[ \bar h\,S \big]\,\Gamma_j\,(\nslash\,\calAslash_{c\perp})\,
    \big[ S^\dagger q_s \big] \\
   \mbox{or} \quad
   &\big[ \bar\xi\,W \big]\,\Gamma_i'\,(\nslash\,\calAslash_{c\perp})\,
    \big[ S^\dagger q_s \big]\,
    \big[ \bar h\,S \big]\,\Gamma_j'\,\big[ W^\dagger\xi \big] \,,
\end{aligned}
\end{equation}
where each bracket $[\dots]$ can be located at a different point, and the 
factor $(\nslash\,\calAslash_{c\perp})$ may or may not be present. Note 
that the second operator consists of a product of a soft-collinear 
current considered in Section~\ref{sec:softcol} with a heavy-collinear 
current derived in \cite{Bauer:2000yr}. With these results at hand, one 
can now perform a Fierz transformation to the flavor basis 
$\bar\xi\dots\xi\,\bar h\dots q_s$ and express the result in terms of 
color singlet-singlet and color octet-octet operators, as shown in 
(\ref{4qfinal}).

Consider now the hadronic matrix elements of the resulting SCET operators 
between an initial-state $B$ meson and a light, highly energetic 
final-state meson $M$. Since at leading order in $\lambda$ there are no 
QCD interactions between soft and collinear fields, it follows that these
matrix elements factorize into two parts, one containing only soft fields
and one containing only collinear fields. Using the fact that matrix 
elements of color-octet currents between physical states vanish, and 
recalling from (\ref{Adef}) that the object $\A_c^\mu$ is a pure color 
octet, we obtain
\begin{eqnarray}
   \langle M|\,Q_{ij}^{(S)}(r,t)\,|B\rangle
   &=& \langle M|\,\big[ \bar\xi\,W \big](-r\bar n)\,\Gamma_i\,
    \big[ W^\dagger\xi \big](r\bar n)\,|\,0\,\rangle\,
    \langle\,0\,|\,\big[ \bar h\,S \big](0)\,\Gamma_j\,
    \big[ S^\dagger q_s \big](tn)\,|B\rangle \nonumber\\
   \langle M|\,R_{ij}^{(O)}(r,s,t)\,|B\rangle
   &=& \frac{1}{2N_c}\,\langle M|\,\big[ \bar\xi\,W \big](-r\bar n)\,
    \Gamma_i\,\A_c^\mu(s\bar n)\,
    \big[ W^\dagger\xi \big](r\bar n)\,|\,0\,\rangle
    \nonumber\\[-0.2cm]
   &&\hspace{0.5cm}\times 
    \langle\,0\,|\,\big[ \bar h\,S \big](0)\,\Gamma_j\,\nslash\,
    \gamma_{\perp\mu}\,\big[ S^\dagger q_s \big](tn)\,|B\rangle \,,
\end{eqnarray}
while
\begin{equation}
   \langle M|\,Q_{ij}^{(O)}(r,t)\,|B\rangle = 0 \,, \qquad
   \langle M|\,R_{ij}^{(S)}(r,s,t)\,|B\rangle = 0 \,.
\end{equation}
The matrix elements of the singlet operators $Q_{ij}^{(S)}$ have 
precisely the form expected from familiar applications of QCD 
factorization. The two matrix elements of bilocal currents define the 
leading-twist light-cone distribution amplitude of the light meson $M$
 \cite{braun,Ball:1998sk} and the leading-order (in the heavy-quark 
expansion) distribution amplitudes of the $B$ meson \cite{Grozin:1996pq}, 
respectively. The fact that there may appear a non-vanishing, 
leading-power contribution from matrix elements of the octet operators 
$R_{ij}^{(O)}$ is a surprising result of our analysis, which has not been 
anticipated in the literature. These matrix elements correspond to 
higher-twist projections onto the light meson $M$, which contribute at 
leading power because (in momentum space) they are enhanced with respect 
to the singlet-operator matrix elements by a factor of $1/l_+$, where $l$ 
is the soft spectator momentum. Introducing the definition 
$W(x,y)\equiv W(x)\,W^\dagger(y)$ for a collinear Wilson line connecting 
two points $x$ and $y$, and using relation (\ref{Adef}), the 
corresponding matrix element can be recast into the form
\begin{eqnarray}
   &&\langle M|\,\big[ \bar\xi\,W \big](-r\bar n)\,\Gamma_i\,
    \A_c^\mu(s\bar n)\,\big[ W^\dagger\xi \big](r\bar n)\,|\,0\rangle
    \nonumber\\
   &&= \int_{-\infty}^s\!dw\,\langle M|\,\bar\xi(-r\bar n)\,
    W(-r\bar n,w\bar n)\,\Gamma_i\,\bar n_\alpha\,
    g G_c^{\alpha\mu}(w\bar n)\,W(w\bar n,r\bar n)\,\xi(r\bar n)\,
    |\,0\,\rangle \,, \qquad
\end{eqnarray}
which involves the conventional definition of a higher-twist, 
three-particle light-cone distribution amplitude 
\cite{braun,Ball:1998sk}. It remains to be seen whether such higher
Fock-state contributions will also arise in cases where a non-local 
amplitude is matched onto four-quark operators in the SCET. 

The momentum-space Wilson coefficients corresponding to the coefficients
$\widetilde C_{ij}^{(C)}$ and $\widetilde D_{ij}^{(C)}$ in 
(\ref{4qfinal}) are, in general, complicated functions of the scales 
$m_b^2$, $m_b E$, $E l_+$, and $\mu^2$ (where $E\sim m_b$), and of 
dimensionless variables $u$ and $v$ measuring the longitudinal momentum 
fractions of the collinear quark and gluon (in the case of the operators 
$R_{ij}^{(C)}$) inside the light final-state meson. It is impossible to 
eliminate all large ratios of these scales by a choice of the 
renormalization point $\mu$. In such a case the resummation of large 
logarithms of the type $\ln(E/\Lambda)$ can be achieved by performing the 
matching onto the low-energy effective theory in two steps. First one 
integrates out hard modes as well as the couplings of collinear gluons to 
the heavy quark. This yields (non-local) operators in an intermediate 
effective theory which still contains $(\lambda,1,\lambda)$ modes as 
dynamical degrees of freedom. The Wilson coefficients arising in this 
step are functions of $m_b^2$, $m_b E$ and $\mu^2$ (as well as $u$). They 
can be calculated perturbatively at a scale $\mu^2\sim m_b^2$ and evolved 
down to a scale $\mu^2\sim E\Lambda$ using the renormalization group. In 
the second step the off-shell $(\lambda,1,\lambda)$ modes are integrated 
out, yielding the SCET as constructed in this work. This gives rise to 
Wilson coefficients that depend on the scales $E l_+$ and $\mu^2$ (as 
well as $u$ and $v$). Solving the renormalization-group equations in SCET 
these coefficients can then be evolved down to scales 
$\mu^2\ll E\Lambda$, at which the operators in SCET are renormalized. 
Concrete examples of such a two-step matching procedure will be discussed 
elsewhere. We stress that, while the resummation of large logarithms 
arising from the evolution between the two hard scales $m_b^2$ and 
$m_b\Lambda$ is necessary to obtain reliable perturbative predictions for 
the Wilson coefficient functions, it is not required for the discussion 
of the factorization properties of matrix elements in the low-energy 
effective theory.

\section{Implications for factorization theorems}
\label{sec:examples}

The presence of non-trivial interactions between the soft spectator 
quark and collinear gluons complicates the understanding of the 
factorization properties of $B$-meson decay amplitudes. We will now 
illustrate this fact with the help of some toy examples. Realistic 
examples such as $B\to K^*\gamma$ or $B\to\pi\pi$ are more complicated 
and will be discussed elsewhere.

Consider the effective weak Hamiltonian 
\begin{equation}\label{Heff}
   {\cal H}_{\rm eff}
   = C^{(S)}\,\Phi\,\bar u\,\Gamma_1\,s\,\bar b\,\Gamma_2\,u
   + C^{(O)}\,\Phi\,\bar u\,\Gamma_1\,t_a\,s\,\bar b\,\Gamma_2\,t_a\,u
\end{equation}
mediating the exclusive decay $B^+\to K^{(*)+}\Phi$, where the 
color-singlet field $\Phi$ can be a scalar, vector, or any other field, 
depending on the quantum numbers of the Dirac matrices $\Gamma_{1,2}$. 
This interaction is simpler than in the Standard Model since it is local, 
but otherwise it shares many similarities with terms that occur in the 
effective weak Hamiltonian of the Standard Model.

Based on the fact that the outgoing kaon contains a highly energetic, 
collinear $(\bar s u)$ pair and so decouples from soft gluons (``color
transparency''), one would expect, following 
\cite{Beneke:1999br,Beneke:2001ev}, that at leading power in 
$\Lambda/m_b$ the decay amplitude should obey the QCD factorization 
formula
\begin{equation}\label{ff}
   {\cal A} = \Phi \int du\int dl_+\,\phi_{K^{(*)}}(u)\,\phi_B(l_+)\,
   T(u,l_+) \,,
\end{equation}
where $\phi_{K^{(*)}}$ and $\phi_B$ are leading-order light-cone 
distribution amplitudes defined, e.g., in \cite{Beneke:2000ry}, and $T$ 
is the hard-scattering kernel, which in the context of the SCET would be
identified with a Wilson coefficient function. It is evident from our 
discussion in the previous section that for the simple result (\ref{ff}) 
to be correct the matrix elements of the operators $R_{ij}^{(C)}$ would 
have to vanish. Otherwise, the factorization formula would have to be 
generalized to include a term involving higher-twist distribution 
amplitudes of the kaon.

Matrix elements of the operators $R_{ij}^{(C)}$ in (\ref{QRdef}) are zero 
if the product $\Gamma_j\,\nslash$ of Dirac matrices vanishes, or if the 
$B$-meson matrix element of the structure 
$\bar h\,\Gamma_j\,\nslash\,\gamma_{\perp\mu}\,q_s$ vanishes by 
rotational invariance in the transverse plane. We believe that in many
cases of phenomenological interest this is indeed what happens. Consider, 
as an example, the decay $B^+\to K^+\Phi^0$, where $\Phi^0$ is a 
fictitious light scalar. In this case the Lorentz indices of $\Gamma_1$ 
must always be contracted with those of $\Gamma_2$. Let us, for 
simplicity, work at tree level, so that the Dirac structures 
$\Gamma_i\otimes\Gamma_j$ appearing in the SCET operators are the same as 
those in the original operators. Between the collinear quark spinors 
$\bar\xi$ and $\xi$ the Dirac basis matrices 
$\{1,\gamma_5,\gamma^\mu,\gamma^\mu\gamma_5,[\gamma^\mu,\gamma^\nu]\}$ 
are projected to $\{0,0,\frac12 n^\mu\,\nbslash,
\frac12 n^\mu\,\nbslash\gamma_5,\nbslash(n^\mu\gamma_\perp^\nu-
n^\nu\gamma_\perp^\mu)\}$. It follows that after contraction of Lorentz
indices the product $\Gamma_i\otimes\Gamma_j$ can only take the forms
$\nbslash(\gamma_5)\otimes\nslash(\gamma_5)$ or
$[\nbslash,\gamma_\perp^\nu]\otimes[\nslash,\gamma_{\perp\nu}]$. This
guarantees that $\Gamma_j\,\nslash=0$ for all possible Dirac structures, 
and hence the operators $R_{ij}^{(C)}$ vanish. We conclude that for this 
particular process the factorization formula (\ref{ff}) would hold, at 
least at tree level.

The mechanism just described does not appear to be universal, however. 
Consider, as a counterexample, the case where $\Phi=F_{\alpha\beta}$ is 
the electromagnetic field, and where the Dirac matrices in (\ref{Heff}) 
are chosen to be 
$\Gamma_1\otimes\Gamma_2=\sigma^{\alpha\beta}\otimes\gamma_5$ for the 
color singlet-singlet term and 
$\gamma^\alpha\otimes\gamma^\beta\gamma_5$ for the octet-octet term. 
This effective Hamiltonian mediates the radiative decay 
$B^+\to K^{*+}\gamma$ (but not in the Standard Model). At tree level, the 
corresponding operators $Q_{ij}^{(C)}$ and $R_{ij}^{(C)}$ in SCET have 
Dirac structures 
$\Gamma_i\otimes\Gamma_j=\nbslash\,\gamma_\perp^\beta\otimes\gamma_5$ 
times $n^\alpha F_{\alpha\beta}$ and
$\Gamma_i\otimes\Gamma_j\,\nslash\,\gamma_{\perp\mu}
=\nbslash\otimes\gamma^\beta\gamma_5\,\nslash\,\gamma_{\perp\mu}$ times 
$n^\alpha F_{\alpha\beta}$, respectively. After projection onto the $B$ 
meson, the resulting Lorentz structures in the two cases are
\begin{equation}
\begin{aligned}
   \int\!dr\,dt\,\widetilde C_{ij}^{(S)}\,
   \langle K^*|\,Q_{ij}^{(S)}\,|B\rangle
   &\sim E^{\frac12}\Lambda^{\frac32}\,n^\alpha F_{\alpha\beta}\,
    \langle K^*|\,\bar\xi\,\nbslash\,\gamma_\perp^\beta\,\xi\,
    |\,0\,\rangle
    \sim E^{\frac32}\Lambda^{\frac52}\,n^\alpha\,
    \varepsilon_\perp^{*\beta} F_{\alpha\beta} \,, \\
   \int\!dr\,ds\,dt\,\widetilde D_{ij}^{(O)}\,
   \langle K^*|\,R_{ij}^{(O)}\,|B\rangle
   &\sim E^{\frac12}\Lambda^{\frac12}\,n^\alpha F_{\alpha\beta}\,
    g_{\perp\mu}^\beta\,\langle K^*|\,
    \bar\xi\,\nbslash\,\A_{c\perp}^\mu\,\xi\,|\,0\,\rangle
    \sim E^{\frac32}\Lambda^{\frac52}\,n^\alpha\,
    \varepsilon_\perp^{*\beta} F_{\alpha\beta} \,,
\end{aligned}
\end{equation}
where $\varepsilon_\perp$ is the transverse polarization vector of the 
$K^*$ meson, and we have used the well-known scaling relations for 
current matrix elements of heavy and light mesons \cite{Neubert:1993mb}. 
It is evident that for this example both operators contribute at the same 
order in power counting, and hence the simple QCD factorization formula 
in (\ref{ff}) must be generalized to include a term involving twist-3 
light-cone distribution amplitudes of the kaon.

It follows that the question of factorization for the hard-spectator term 
in a QCD factorization formula is far from trivial. Whether a simple QCD 
factorization formula holds, or whether it must be generalized to include 
higher-twist distribution amplitudes, requires a case by case study. It 
is conceivable that in many cases of phenomenological importance it will 
be possible to exclude the presence of the operators $R_{ij}^{(C)}$ based 
on some symmetry, such as rotational invariance in the transverse plane 
or reparameterization invariance. It remains to find a general argument 
supporting this assertion.

\section{Gauge-invariant building blocks for operators in SCET}
\label{sec:pretty}

The careful reader will have noticed that, as a result of gauge 
invariance, the external current and four-quark operators in SCET 
discussed in the previous sections always contain products of the Wilson 
lines $W$ and $S$ with soft or collinear quark fields, respectively. This 
suggests introducing new fields
\begin{equation}
   \H = S^\dagger h \,, \qquad 
   \Q_s = S^\dagger q_s \,, \qquad 
   \X = W^\dagger\xi \,,
\end{equation}
which are manifestly gauge invariant and have the same scaling relations 
in $\lambda$ as the original fields. In the case where one chooses the
light-cone gauge conditions $n\cdot A_s=0$ and $\bar n\cdot A_c=0$ the
new fields agree with the original ones. In terms of the new fields, the 
four-quark operators in (\ref{QRdef}) take the simple form
\begin{equation}
\begin{aligned}
   Q_{ij}^{(C)}(r,t) &= \bar\X(-r\bar n)\,\Gamma_i\,T_1\,\X(r\bar n)\,
    \bar\H(0)\,\Gamma_j\,T_2\,\Q_s(tn) \,, \\
   R_{ij}^{(C)}(r,s,t) &= \bar\X(-r\bar n)\,\Gamma_i\,T_1\,
    \X(r\bar n)\,\bar\H(0)\,\Gamma_j\,T_2\,\nslash\,
    \calAslash_{c\perp}(s\bar n)\,\Q_s(tn) \,,
\end{aligned}
\end{equation}
and similar simplified expressions hold for the current operators in 
(\ref{HLmatch}), (\ref{LLmatch1}), and (\ref{LLmatch2}). 

While the definition of these new fields seems merely a matter of 
convenience of notation, we will now argue that, in fact, any operator 
in SCET can be built from a small set of gauge-invariant fields. Once 
these fields are known, gauge invariance of operators containing them is 
guaranteed, and the only remaining rules for the construction of SCET 
operators are Lorentz and reparameterization invariance. Therefore, the 
formalism we will introduce in this section will simplify the 
construction of SCET operators, which is important, in particular, for 
going beyond the leading order in $\lambda$.

The QCD Lagrangian contains, besides quark (and ghost) fields, the 
gauge-covariant derivative. In SCET we have distinguished between the 
collinear and soft covariant derivatives $D_c^\mu$ and $D_s^\mu$. We can 
construct gauge-invariant objects from these quantities by using the 
field $\A_c^\mu$ introduced in (\ref{Adef}) and a corresponding object 
defined in the soft sector:
\begin{equation}
   \A_s^\mu(x) = \big[ S^\dagger(iD_s^\mu S) \big](x)
   = \int_{-\infty}^0\!dw\,n_\alpha\,\big[ S^\dagger g G_s^{\alpha\mu}
   S \big](x+wn) \,.
\end{equation}
The gauge-invariant fields $\A_c^\mu$ and $\A_s^\mu$ obey the 
constraints $\bar n\cdot\A_c=0$ and $n\cdot\A_s=0$ in any gauge. (Recall 
that in the light-cone gauge these fields coincide with the corresponding 
gluon fields $g A_c^\mu$ and $g A_s^\mu$.) It follows that the scaling 
relations of the new fields are $\A_c^\mu\sim(\lambda^2,0,\lambda)$ and 
$\A_s^\mu\sim(0,\lambda,\lambda)$. We finally define new, gauge-invariant 
objects
\begin{equation}
   i\D_c^\mu = W^\dagger\,iD_c^\mu\,W = i\partial^\mu + \A_c^\mu \,,
   \qquad
   i\D_s^\mu = S^\dagger\,iD_s^\mu\,S = i\partial^\mu + \A_s^\mu \,.
\end{equation}

With the help of these definitions, the different parts of the 
leading-order SCET Lagrangian can be rewritten in the form
\begin{equation}\label{Lrewrite}
\begin{aligned}
   {\cal L}_h &= \bar\H\,iv\cdot\D_s\,\H \,, \qquad
    {\cal L}_s = \bar\Q_s\,i\calDslash_s\,\Q_s \,, \\
   {\cal L}_c &= \bar\X\,\frac{\nbslash}{2}\,in\cdot\D_c\,\X 
    + \bar\X\,\frac{\nbslash}{2}\,i\calDslash_{c\perp}\,
    \frac{1}{i\bar n\cdot\partial}\,i\calDslash_{c\perp}\,\X \,.
\end{aligned}
\end{equation}
The appearance of an ordinary derivative in the expression for 
${\cal L}_c$ is not in conflict with gauge invariance, since all the 
component fields are gauge invariant by themselves. Finally, also the 
pure-glue Lagrangians for soft and collinear fields take a simple form. 
Since we can replace $G_c^{\mu\nu}$ by 
$W^\dagger G_c^{\mu\nu}\,W=(i/g)\,[\D_c^\mu,\D_c^\nu]$ and $G_s^{\mu\nu}$ 
by $S^\dagger G_s^{\mu\nu} S=(i/g)\,[\D_s^\mu,\D_s^\nu]$ in the 
Lagrangian (using the cyclicity of the color trace), the resulting 
pure-glue Lagrangians are simply obtained by replacing all gluon fields 
in the usual QCD Lagrangian with the new fields $\A_c$ and $\A_s$.

Let us briefly summarize the properties of the gauge-invariant building 
blocks under the three types of reparameterizations discussed in 
Section~\ref{sec:RPI}. We find that the redefined quark fields as well as 
the soft gluon field $\A_s^\mu$ are invariant up to higher-order 
terms in $\lambda$. However, the collinear gluon field has non-trivial 
transformations at leading order. Under a type~I transformation
$n\cdot\A_c\to n\cdot\A_c+\epsilon_\perp\cdot\A_c$ while 
$\A_{c\perp}$ remains invariant. Under a type~II transformation 
$\A_{c\perp}^\mu\to 
\A_{c\perp}^\mu-\frac12\,n^\mu\,e_\perp\cdot\A_{c\perp}$ while 
$n\cdot\A_c$ remains invariant. The corresponding transformation 
properties of the ``collinear derivative'' are
$n\cdot\D_c\to n\cdot\D_c+\epsilon_\perp\cdot\D_c$ and 
$\D_{c\perp}^\mu\to \D_{c\perp}^\mu -\frac12\,\epsilon_\perp^\mu\,
\bar n\cdot\partial$ (type~I), and $\D_{c\perp}^\mu\to \D_{c\perp}^\mu
-\frac12\,n^\mu\,e_\perp\cdot\D_{c\perp}$ while $n\cdot\D_c$ remains 
invariant (type~II). Invariance under type~III reparameterizations 
requires that every occurrence of a vector $n$ in the numerator must be 
accompanied by that of a vector $\bar n$, or by a factor of 
$n\cdot\partial$ in the denominator (and vice versa). From these rules
it follows, e.g., that the sum of the two terms in the collinear
Lagrangian ${\cal L}_c$ in (\ref{Lrewrite}) is reparameterization 
invariant, but not the two operators separately.

After the introduction of the new fields the expression for any operator 
in SCET looks like an expression in the light-cone gauge; however, we 
have not imposed light-cone gauge but rather redefined the fields by a 
unitary transformation. In the new formulation the Wilson lines have 
disappeared, and all fields in the theory scale like at least one power 
of $\lambda$ (the large collinear field $\bar n\cdot A_c\sim 1$ has been
removed). Finally, gauge invariance is no longer a constraint on the 
construction of operators in the effective theory.

\section{Kinematics of heavy-to-light form factors}
\label{sec:formfactor}

Our goal in this paper was to find a formulation of SCET suitable for the 
systematic study of the factorization properties and power corrections 
for any exclusive $B$-meson decay into light particles. It is important, 
then, to have a theory that describes the form-factor term and the 
hard-scattering term in a factorization formula in terms of the same 
fields and scaling rules. However, previous work on the soft component of 
exclusive heavy-to-light form factors has been based on the scaling 
assumption $\lambda\sim\sqrt{\Lambda/E}$ \cite{Chay:2002vy,Beneke:2002ph},
which is different from our hypothesis. In this case soft fields carrying 
momenta of order $\Lambda$ scale like $E(\lambda^2,\lambda^2,\lambda^2)$
and are thus called ``ultrasoft''. The main difference with our approach
is that the ``collinear modes'' in this formulation have momenta scaling 
like $(\Lambda,E,\sqrt{E\Lambda})$. The resulting effective theory thus 
appears to be different from the one constructed here.

The challenge is to understand the soft contribution to heavy-to-light 
form factors, in which the $B$-meson spectator quark enters the light 
final-state meson as a soft (or ultrasoft) quark, so that this meson is
produced in a highly asymmetric state. Consider a $B\to\pi$ transition
for concreteness. The scaling $\lambda\sim\sqrt{\Lambda/E}$ was assumed 
based on the following kinematical consideration. The pion emitted in a 
heavy-to-light decay at large recoil carries momentum scaling like 
$p_\pi\sim(\Lambda^2/E,E,\Lambda)$, and making this up by combining a 
soft quark with momentum $p_s\sim(\Lambda,\Lambda,\Lambda)$ and a 
collinear jet requires that this jet have invariant mass squared 
$p_c^2\sim E\Lambda$. Given that for a collinear particle 
$p_c^2\sim\lambda^2 E^2$, it then follows that 
$\lambda\sim\sqrt{\Lambda/E}$. Although this argument might seem 
compelling, it has the unattractive feature that the external pion 
momentum now scales like $p_\pi\sim E(\lambda^4,1,\lambda^2)$, which 
cannot be built up from the combination of a generic ultrasoft momentum
$p_s\sim E(\lambda^2,\lambda^2,\lambda^2)$ with a generic collinear 
momentum $p_c\sim E(\lambda^2,1,\lambda)$. In other words, the plus 
component of the collinear jet must cancel the plus component of the soft 
spectator quark with a relative precision of order 
$\lambda^2\sim\Lambda/E$, and the transverse component of the jet has to 
be smaller than its generic size by a factor 
$\lambda\sim\sqrt{\Lambda/E}$. These two constraints combined imply that 
the soft overlap mechanism is strongly suppressed in this picture (which 
a priori is no problem, since heavy-to-light form factors are suppressed 
in the heavy-quark limit). This discussion can also be summarized in more 
physical terms by saying that it is unlikely that a collinear jet of 
particles with invariant mass of order $\sqrt{E\Lambda}$ will absorb a 
soft quark and become a light meson with mass of order $\Lambda$.

\begin{figure}
\epsfxsize=6.0cm
\centerline{\epsffile{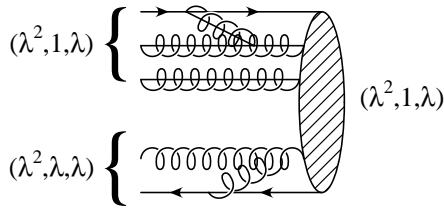}}
\vspace{0.0cm}
\centerline{\parbox{14cm}{\caption{\label{fig:pion}
How to make a pion in a highly asymmetric configuration? The drawing
illustrates the scaling of the collinear and soft components proposed
in the present work, where $\lambda\sim\Lambda/E$ is assumed. In earlier 
papers a different scaling of collinear momenta was used.}}}
\end{figure}

Here we wish to suggest a different possibility for interpreting the soft
overlap contribution to heavy-to-light form factors. As illustrated in
Figure~\ref{fig:pion}, we assume the scaling $\lambda\sim\Lambda/E$ 
adopted throughout this work, so that the pion momentum scales like any 
other collinear momentum. In order to make a light meson out of collinear 
particles and soft particles, the only thing that is required is that the 
plus component of the total soft momentum, which generically would scale 
like $E\lambda$, is accidentally small, of order $E\lambda^2$. This 
implies a phase-space suppression of order $\Lambda/E$. It appears more 
intuitive to us to assume that only a jet of particles with invariant 
mass of order $\Lambda$ can ultimately hadronize into a single light 
meson. 

It is not evident whether the choice of $\lambda$ is simply a matter of 
convenience, or whether it corresponds to a different kinematical 
situation. Naively, we would expect that our predictions for 
heavy-to-light form factors would differ from the ones obtained in 
\cite{Chay:2002vy,Beneke:2002ph}. For instance, we expect that violations 
of heavy-quark symmetry relations between form factors start at order 
$\Lambda/m_b$, while the power counting adopted in these papers would 
allow for the presence of $\sqrt{\Lambda/m_b}$ corrections.

\section{Summary and conclusions}
\label{sec:concl}

The development of an effective field theory for the strong interactions
of soft and collinear partons is a significant step toward the systematic 
study of factorization and a field-theoretical description of power 
corrections for observables that do not admit an operator product 
expansion. Power counting in this soft-collinear effective theory is 
non-trivial due to the presence of non-local operators integrated along 
light-like directions, but it appears feasible to construct a controlled 
heavy-quark expansion of amplitudes in terms of hadronic matrix elements 
of effective theory operators.

In this paper we have constructed the extension of the previous 
formulation of soft-collinear effective theory necessary for the 
description of exclusive $B$-meson decays into light particles. QCD 
factorization theorems for such processes are complicated, because in 
addition to a form-factor term a hard-scattering contribution appears at 
leading power. It results from hard gluon exchange with the soft 
spectator quark in the $B$ meson. One of the main findings of our study 
is that interactions of collinear gluons with soft light quarks are more 
complicated than the corresponding interactions with heavy quarks. In 
particular, transverse collinear gluons have unsuppressed couplings to 
light soft quarks, while only longitudinal collinear gluons couple to 
heavy quarks at leading order in power counting. The intrinsic softness 
of the $B$-meson dynamics complicates the understanding of factorization 
properties of decay amplitudes. A new intermediate mass scale of order 
$m_b\Lambda$ arises and leads to large non-localities of effective-theory 
operators on a scale $1/\Lambda$, thus upsetting naive power counting. 
Power counting can be restored, however, using reparameterization 
invariance, which gives control over the dependence of Wilson coefficient 
functions on the light-like separation between the component fields of 
non-local operators.

The version of soft-collinear effective theory relevant to the discussion
of exclusive $B$ decays into light particles contains soft and collinear
fields. We have constructed the effective Lagrangian at leading order
in $\Lambda/m_b$, finding that it does not contain soft-collinear 
interaction terms. Such interactions would, however, appear at subleading 
order. We have then discussed in detail the matching of current and 
four-quark operators from the full theory onto their effective-theory 
counterparts. This matching is complicated by the presence of non-trivial 
interactions between the soft spectator quark and collinear gluons. The 
most surprising finding of our analysis is that, generically, there are 
two types of four-quark operators present in the effective theory, one of 
which contains an insertion of a transverse collinear gluon field. Upon 
evaluating hadronic matrix elements of such operators between a $B$ meson 
and a light, energetic meson, one finds that in addition to the 
leading-twist distribution amplitude of the light meson also 
three-particle distribution amplitudes of subleading twist can contribute 
at leading power. This suggests that, in some cases, QCD factorization 
formulae may have to be generalized. We have presented a toy example 
where this extension is indeed necessary.

Our results for the matching of currents and four-quark operators suggest 
a reformulation of soft-collinear effective theory in terms of operators 
composed out of gauge-invariant building blocks replacing the original 
quark and gluon fields. In the new formulation gauge invariance is 
automatic, and the form of operators is only constrained by Lorentz 
invariance. We anticipate that this observation will facilitate the
extension of our results beyond the leading order. 

The formalism developed in this work provides for the first time the 
basis for a systematic discussion of factorization and power corrections 
for any $B$-meson decay into light particles. Phenomenological 
applications of this framework will be discussed elsewhere.

\vspace{0.3cm}  
{\it Note added:\/}
While this paper was in writing the work \cite{Lunghi:2002ju} appeared,
in which factorization for the leptonic radiative decay $B\to\gamma e\nu$ 
is discussed in the context of soft-collinear effective theory. While the 
formalism used in that work differs from the one developed here, part of 
the discussion presented by these authors realizes the idea of two-step 
matching mentioned at the end of Section~\ref{sec:4qops}.

\vspace{0.3cm}  
{\it Acknowledgment:\/} 
We are grateful to Martin Beneke, Peter Lepage, Ben Pecjak, Bj\"orn 
Lange, Farrukh Chishtie and Stefan Bosch for many useful discussions. The 
research of R.J.H.\ was supported by the Department of Energy under Grant 
DE-AC03-76SF00515. The research of M.N.\ was supported by the National 
Science Foundation under Grant PHY-0098631.


\begin{thebibliography}{99}

\bibitem{Bauer:2000ew}
C.~W.~Bauer, S.~Fleming and M.~E.~Luke,
Phys.\ Rev.\ D {\bf 63}, 014006 (2001)
[hep-ph/0005275].

\bibitem{Bauer:2000yr}
C.~W.~Bauer, S.~Fleming, D.~Pirjol and I.~W.~Stewart,
Phys.\ Rev.\ D {\bf 63}, 114020 (2001)
[hep-ph/0011336].

\bibitem{Bauer:2001ct}
C.~W.~Bauer and I.~W.~Stewart,
Phys.\ Lett.\ B {\bf 516}, 134 (2001)
[hep-ph/0107001].

\bibitem{Bauer:2001yt}
C.~W.~Bauer, D.~Pirjol and I.~W.~Stewart,
Phys.\ Rev.\ D {\bf 65}, 054022 (2002)
[hep-ph/0109045].

\bibitem{Beneke:2000ry}
M.~Beneke, G.~Buchalla, M.~Neubert and C.~T.~Sachrajda,
Nucl.\ Phys.\ B {\bf 591}, 313 (2000)
[hep-ph/0006124].

\bibitem{Bauer:2001cu}
C.~W.~Bauer, D.~Pirjol and I.~W.~Stewart,
Phys.\ Rev.\ Lett.\ {\bf 87}, 201806 (2001)
[hep-ph/0107002].

\bibitem{Bauer:2002nz}
C.~W.~Bauer, S.~Fleming, D.~Pirjol, I.~Z.~Rothstein and I.~W.~Stewart,
Phys.\ Rev.\ D {\bf 66}, 014017 (2002)
[hep-ph/0202088].

\bibitem{Chay:2002vy}
J.~Chay and C.~Kim,
Phys.\ Rev.\ D {\bf 65}, 114016 (2002)
[hep-ph/0201197].

\bibitem{Beneke:2002ph}
M.~Beneke, A.~P.~Chapovsky, M.~Diehl and T.~Feldmann,
Nucl.\ Phys.\ B {\bf 643}, 431 (2002)
[hep-ph/0206152].

\bibitem{Beneke:1999br}
M.~Beneke, G.~Buchalla, M.~Neubert and C.~T.~Sachrajda,
Phys.\ Rev.\ Lett.\ {\bf 83}, 1914 (1999)
[hep-ph/9905312].

\bibitem{Beneke:2001ev}
M.~Beneke, G.~Buchalla, M.~Neubert and C.~T.~Sachrajda,
Nucl.\ Phys.\ B {\bf 606}, 245 (2001)
[hep-ph/0104110].

\bibitem{Bosch:2001gv}
S.~W.~Bosch and G.~Buchalla,
Nucl.\ Phys.\ B {\bf 621}, 459 (2002)
[hep-ph/0106081].

\bibitem{Beneke:2001at}
M.~Beneke, T.~Feldmann and D.~Seidel,
Nucl.\ Phys.\ B {\bf 612}, 25 (2001)
[hep-ph/0106067].

\bibitem{Beneke:2000wa}
M.~Beneke and T.~Feldmann,
Nucl.\ Phys.\ B {\bf 592}, 3 (2001)
[hep-ph/0008255].

\bibitem{Korchemsky:1999qb}
G.~P.~Korchemsky, D.~Pirjol and T.~M.~Yan,
Phys.\ Rev.\ D {\bf 61}, 114510 (2000)
[hep-ph/9911427].

\bibitem{Bosch:2002bv}
S.~W.~Bosch and G.~Buchalla,
JHEP {\bf 0208}, 054 (2002)
[hep-ph/0208202].

\bibitem{Descotes-Genon:2002mw}
S.~Descotes-Genon and C.~T.~Sachrajda,
Nucl.\ Phys.\ B {\bf 650}, 356 (2003)
[hep-ph/0209216].

\bibitem{Grozin:1996pq}
A.~G.~Grozin and M.~Neubert,
Phys.\ Rev.\ D {\bf 55}, 272 (1997)
[hep-ph/9607366].

\bibitem{braun}
V.~M.~Braun and I.~E.~Filyanov, Z.\ Phys.\ C {\bf 48}, 239 (1990).

\bibitem{Ball:1998sk}
P.~Ball, V.~M.~Braun, Y.~Koike and K.~Tanaka,
Nucl.\ Phys.\ B {\bf 529}, 323 (1998)
[hep-ph/9802299].

\bibitem{Kawamura:2001jm}
H.~Kawamura, J.~Kodaira, C.~F.~Qiao and K.~Tanaka,
Phys.\ Lett.\ B {\bf 523}, 111 (2001)
[Erratum: {\em ibid.} B {\bf 536}, 344 (2002)]
[hep-ph/0109181].

\bibitem{Neubert:1993mb}
For a review, see:
M.~Neubert,
Phys.\ Rept.\ {\bf 245}, 259 (1994)
[hep-ph/9306320].

\bibitem{Coleman}
S. Coleman and R. Norton, Nuov.\ Cim.\ {\bf 38}, 438 (1965).

\bibitem{Manohar:2002fd}
A.~V.~Manohar, T.~Mehen, D.~Pirjol and I.~W.~Stewart,
Phys.\ Lett.\ B {\bf 539}, 59 (2002)
[hep-ph/0204229].

\bibitem{Eichten:1990vp}
E.~Eichten and B.~Hill,
Phys.\ Lett.\ B {\bf 243}, 427 (1990).

\bibitem{Neubert:1993za}
M.~Neubert,
Phys.\ Rev.\ D {\bf 49}, 1542 (1994)
[hep-ph/9308369].

\bibitem{Lunghi:2002ju}
E.~Lunghi, D.~Pirjol and D.~Wyler,
Nucl.\ Phys.\ B {\bf 649}, 349 (2003)
[hep-ph/0210091].

\end{thebibliography}
\end{document}